\documentclass[letterpaper,12pt]{article}

\usepackage[T1]{fontenc}
\usepackage{libertine}
\usepackage{caption}
\usepackage[normalem]{ulem}
\usepackage{comment}
\usepackage{url}
\usepackage{tabularx} 
\usepackage{multicol}
\usepackage{multirow}
\usepackage[linkcolor=blue,citecolor=blue]{hyperref}
\usepackage{caption}
\usepackage{subcaption}
\usepackage{setspace}
\usepackage{booktabs}
\usepackage{float}
\usepackage{xcolor}
\usepackage{bm}
\usepackage{array}
\usepackage{dcolumn}
\usepackage{graphicx}
\usepackage{amsmath}
\usepackage{amsfonts}
\usepackage{amsopn}
\usepackage{amsthm}
\usepackage{lscape}
\usepackage{upgreek}
\usepackage{titlesec}
\usepackage{footnote}

\usepackage[linesnumbered,ruled,vlined]{algorithm2e}

\SetCommentSty{mycommfont}
\SetKwInput{KwInput}{Input}               
\SetKwInput{KwOutput}{Output}

\theoremstyle{definition}

\theoremstyle{remark}
\newtheorem{rmk}{Remark}

\titleformat{\subsection}[runin]{\normalfont\bfseries}{\thesubsection.}{3pt}{}
\titleformat{\subsubsection}[runin]{\normalfont\itshape}{\thesubsubsection.}{3pt}{}

\newcounter{rst}

\usepackage[backend=bibtex, style=nature, citestyle=nature]{biblatex}
\begin{document}

\begin{center}
{Study Duration Prediction for Clinical Trials with Time-to-Event Endpoints Using Mixture Distributions Accounting for Heterogeneous Population}
\end{center}
\begin{center}
Hong Zhang$^{1*}$, Jie Pu$^{1}$, Shibing Deng$^{1}$, Satrajit Roychoudhury$^{1}$, Haitao Chu$^{1, 2}$, Douglas Robinson$^{1}$
\end{center}

{\footnotesize
\begin{flushleft}
$^1$ Global Biometrics and Data Management, Pfizer Inc., New York, New York, USA\\
$^2$ Division of Biostatistics, School of Public Health, University of Minnesota, Minneapolis, Minnesota, USA\\
$*$ Corresponding author: hong.zhang3@pfizer.com
\end{flushleft}
}

\centerline{\bf Abstract}
In the era of precision medicine, more and more clinical trials are now driven or guided by biomarkers, which are patient characteristics objectively measured and evaluated as indicators of normal biological processes, pathogenic processes, or pharmacologic responses to therapeutic interventions. With the overarching objective to optimize and personalize disease management, biomarker-guided clinical trials increase the efficiency by appropriately utilizing prognostic or predictive biomarkers in the design. However, the efficiency gain is often not quantitatively compared to the traditional all-comers design, in which a faster enrollment rate is expected (e.g. due to no restriction to biomarker positive patients) potentially leading to a shorter duration. To accurately predict biomarker-guided trial duration, we propose a general framework using mixture distributions accounting for heterogeneous population. Extensive simulations are performed to evaluate the impact of  heterogeneous population and the dynamics of biomarker characteristics and disease on the study duration. Several influential parameters including median survival time, enrollment rate, biomarker prevalence and effect size are identitied. Re-assessments of two publicly available trials are conducted to empirically validate the prediction accuracy and to demonstrate the practical utility. 
The R package \emph{detest} is developed to implement the proposed method and is publicly available on CRAN.


\section{Introduction}\label{sec1}

In therapeutic areas where the primary outcome is often time to an event of interest, e.g., time to disease progression or death, the statistical power and sample size calculation of a clinical trial is mostly driven by the number of events observed and the statistical analyses are typically carried out after the pre-specified event milestones have occurred. Given the considerable amount of operational efforts and expenses for conducting a clinical trial, it is of great value to accurately predict the study duration, i.e. when the milestone(s) will be reached, for both resource and strategic planning purposes. 

Several parametric and non-parametric models have been proposed to predict trial duration based on factors such as patient enrollment rate and event rate. For example, Rubinstein et al. \cite{rubinstein1981planning} used the Poisson accrual process and exponential survival model to predict trial duration needed for achieving the desired number of events. Bagiella and Heitjan \cite{bagiella2001predicting} developed a Bayesian parametric model to estimate both the point and the interval predictions of milestone time based on the accumulating data from the trial while Ying et al. \cite{ying2004nonparametric} later proposed a non-parametric approach to make the point and interval prediction. In Anisimov's work \cite{anisimov2011predictive}, patient enrollment was modeled using a delayed Poisson process to handle different enrollment rates across centers. To facilitate an easier examination of the relationship between sample size (planned number of participants) and the expected study duration and to assess the variation in study duration for a given sample size, Machida et al. \cite{machida2021} developed a graphical approach and a probability density function of the study duration. 
We refer the interested readers to \cite{heitjan2015real, machida2021} for more comprehensive literature review of study duration prediction methodology.

Under the presence of biomarker subgroups, however, the study duration is dependent on the characteristics of the biomarker. Without accounting for this information, most of the previous work may lead to inaccurate study duration prediction, as evidenced by our empirical results in Figure~\ref{fig3a}. The objective of this article is to introduce a statistical model for predicting study duration under patient heterogeneity. The proposed model is an extension of the work by Machida et al. \cite{machida2021} with a newly added component to incorporate different patient risk subgroups.  This generalization is motivated by the practical importance to inform the design of a prognostic enrichment study that aims at increasing trial efficiency for drug development. For event-driven trials, inclusion of patients with lower risk of developing an outcome related event may limit the chance to detect a clinically meaningful treatment effect and may impact the trial efficiency due to increased sample size or prolonged study duration to achieve the targeted event milestone(s) \cite{us2019enrichment} \cite{freidlin2014biomarker} \cite{jering2022improving}. By increasing the proportion of patients with higher risk, a prognostic enrichment study design may potentially help improve the trial efficiency, e.g., reducing sample size or shortening study duration, compared to an all-comer design. Given the complexity of predicting study duration described earlier, a systematic comparison between an enrichment vs. an all-comer design in trial efficiency is needed to inform decision making during the planning stage. 

The remainder of this article is structured as follows: In Section~\ref{sect:framework}, we present the model framework for calculating trial duration and conduct a theoretical analysis of how various parameters affect the duration. Section~\ref{sect:numerical} is dedicated to evaluating the performance of our proposed calculation methods and showcasing some intriguing applications. In Section~\ref{sect:realdata}, we apply our calculation method to actual trial data to validate its accuracy and demonstrate its practical usefulness in predicting trial duration. Finally, we conclude this paper with some discussions and possible future research directions in Section~\ref{sect:discussion}.

\section{Theoretical Framework for Study Duration\label{sect:framework}} 
Consider a clinical trial of $K$ treatment arms on a patient population with $L$ subgroups. Assume the proportion of patients who are in subgroup $l$ and receive treatment $k$ is $r_{kl}\geq 0$, $\sum_{k=1}^K\sum_{l=1}^L r_{kl}=1$. In the randomized setting where the treatment assignment is independent from the subgroups, we may write $r_{kl}=p_kq_l$, where  $p_k$ is the proportion of treatment arm $k$, and $q_l$ is the prevalence of subgroup $l$.   

Let $T_{kl}$ be the random variable of the time from study initiation to event observed 
for patients who are in subgroup $l$ and receive treatment $k$, 
\begin{align}
\label{equ:Tkl}
T_{kl} = 
\left\{\!
\begin{aligned}
  &+\infty, && V_{kl}>W_{kl}\\
  &U_{kl} + V_{kl}, && V_{kl}\leq W_{kl}
\end{aligned}
\right.
\end{align}
where $U_{kl}$ is the time from study initiation to enrollment, $0<U_{kl}<a$ for some enrollment finishing at a given time $a$, $V_{kl}$ is the time from enrollment to an event, $W_{kl}$ is the time from enrollment to drop-out. $U_{kl}$, $V_{kl}$ and $W_{kl}$ are assumed to be independent. The above definition says when an event is observed, the time to the observed event is the summation of time from study initiation to enrollment and the time from enrollment until that patient experiences an event. However, if a patient is censored, the event is never observed and thus the time to the observed event is defined as infinity. 

To derive the cumulative distribution function (CDF) of $T_{kl}$, we use the total probability theorem to obtain the following integral expression (see Appendix~\ref{sect:proof:equ:FTkl}). 
\begin{align}
\label{equ:FTkl}
F_{T_{kl}}(t) = P(T_{kl}\leq t) = \int_0^{\min\{t, a\}}\int_{0}^{t-u}\int_{v}^{+\infty} f_{W_{kl}}(w)f_{V_{kl}}(v)f_{U_{kl}}(u)dwdvdu, \quad{} 0<t<+\infty,
\end{align}
where $f_{U_{kl}}$, $f_{V_{kl}}$, and $f_{W_{kl}}$ are the probability density functions (PDF) of the enrollment time, event time and drop-out time, respectively. It is straightforward to check that, as $t$ goes to infinity, the probability of observing an event before time $t$ converges to the probability of no censoring. That is,
$\lim_{t\to +\infty}F_{T_{kl}}(t) = P(V_{kl}\leq W_{kl})$. 
We further define $F_{T_{kl}}(+\infty)=1$ to make it a valid CDF. 

Denote $T$ the time from study initiation to an observed event for all patients in the study. $T$ is the mixture of $T_{kl}$, $k=1,...,K$, $l=1,...,L$, with mixing parameters $r_{kl}$. The CDF of $T$ may be written as 
\begin{align}
\label{equ:T}
F_T(t) = \sum_{k=1}^K\sum_{l=1}^L r_{kl}F_{T_{kl}}(t).
\end{align}
We can also check that $\lim_{t\to +\infty}F_{T}(t) = \sum_{k=1}^K\sum_{l=1}^L r_{kl}P(V_{kl}\leq W_{kl})$, and similarly define $F_T(+\infty)=1$.  

\subsection{Study Duration of Clinical Trials with Time-to-Event Endpoints \label{sect:2.1}}
\quad{}\newline 
In clinical trials with time-to-event endpoints, the study planning is often driven by a target number of events, $d$. There are different methods to calculate $d$, such as Schoenfeld formula \cite{schoenfeld1981} and Freedman formula \cite{freedman1982tables}. Here, we assume such $d$ has been determined. Therefore, the study duration may be defined as the time from study initiation to observing the $d$th event, 
\begin{align}
    \text{Study Duration} = T_{(d)}, 
\end{align}
the $d$th order statistics from $T_{(1)}\leq T_{(2)}\leq...\leq T_{(n)}$, where $n$ is the sample size. 

It is important to note that the study duration $T_{(d)}$ is a random variable. In the case of heavy censoring, the $d$th event may never be observed, i.e., $T_{(d)}$ has a non-zero probability of being $+\infty$. However, recognizing that the study duration is random, we may still need a point estimate to guide clinical trial design. One approach is by noticing that, according to the strong law of large numbers for order statistics, 
\begin{align}
\label{equ:Td_est1}
    \lim_{\substack{n\to\infty,\\d/n\to s}}T_{(d)}\to F_T^{-1}(s) \text{ in probability},
\end{align}
where $0<s<+\infty$ is the convergent value for $d/n$. Under large samples, $s\approx d/n$, $F_T^{-1}(d/n)$ may be a good estimate of the study duration. When $F_T(t)$ is known, the percentiles may be efficiently found using the bisection method. 

The above asymptotics may no longer hold, however, under moderate or small sample sizes. In such cases, we may use the median, $med(T_{(d)})$, as our study duration estimate. The distribution of $T_{(d)}$ is theoretically known by the classic order statistics theory \cite{david2004order} when $F_T(t)$ is available. To obtain the median, a direct calculation approach is viable following the similar idea in Machida et al \cite{machida2021}. Alternatively, we propose a simulation based approach (See Algorithm \ref{alg:duration}), which is computationally easier and more stable than the direct calculation approach. It is worth mentioning that both the percentile approach and the median approach may cover any continuous survival and enrollment distributions in \ref{equ:Tkl}. 

\begin{algorithm}[h]
\DontPrintSemicolon
  \KwInput{\setstretch{1.35} Sample size $n$, number of events $d$, number of treatments $K$, proportion of treatment arm $k$, $p_k$, number of biomarker groups $L$, prevalence of biomarker subgroup $l$, $q_l$, CDF of time from study initiation to event observed $F_{T_{kl}}$, number of repetitions $R$ and confidence level $\alpha$.}
  \KwOutput{Study duration estimates.}
  Calculate $P_k=\sum_{i=1}^kp_i, Q_l=\sum_{i=1}^lq_i$, $k=1,...,K$, $l=1,...,L$.\\
  \For{i = 1 to R}    
  {
      \For{j = 1 to n}
      {
          $u \gets$ runif(0, 1).\\
          $v \gets$ runif(0, 1).\\
          \If{$P_{k-1}<u\leq P_k$ $\And$ $Q_{l-1}<v\leq Q_l$}
          {
            generate a random number $x_j$ from $F_{T_{kl}}$, e.g., by inverse transform sampling.
          }

      }
      Find the $d$th order statistic, $x_{(d)}$, from $x_1,..., x_j,..., x_n$. 
  }
  Output the median, $\alpha/2$ th and $1-\alpha/2$ th percentiles of $x_{(d), 1},..., x_{(d), i},... ,x_{(d), R}$.
\caption{Pseudo algorithm for study duration calculation}
\label{alg:duration}
\end{algorithm}

\subsection{Exponential Event Time and Potentially Non-uniform Enrollment \label{sect:nuniform}}
\quad{}\newline 
For some parametric family of survival distributions and enrollment time distribution, we may develop an analytical expression for $F_{kl}$ in equation \ref{equ:FTkl}. Here, we drop the subscript $kl$ to simplify the notations. Assume exponential survival $V\sim Exp(\lambda_V)$, exponential drop-out $W\sim Exp(\lambda_W)$ and enrollment follows a scaled Beta distribution  $U/a\sim Beta(\alpha=1, \beta)$, where $a$ is the enrollment completion time. Its PDF is
\begin{align}
\label{equ:beta}
f_{U}(u) &= \frac{\beta}{a} \left(1-\frac{u}{a}\right)^{\beta-1}, 0< u < a.
\end{align}
This family covers a wide variety of situations for enrollment. When $\beta=1$, the enrollment is uniform over $(0, a)$. When $\beta<1$, the enrollment curve is convex such that the enrollment is slower at the beginning and  then increasingly faster later on. When $\beta>1$, the enrollment curve is concave meaning the enrollment is faster at the beginning and then slows down later.%, which is less common in reality. 

Under these parametric assumptions, we may evaluate (Appendix \ref{sect:proof:equ:nunif}) the integral in equation (\ref{equ:FTkl}) to get the CDF
\begin{align}
\label{equ:nunif}
\begin{split}
F_{T_{kl}}(t) = \frac{\lambda_V}{\lambda_V+\lambda_W}
\left(1 - \max\{0, \frac{a-t}{a}\}^\beta - \frac{\beta\Gamma(\beta)e^{-(\lambda_V+\lambda_W)(t-a)}}{(a(\lambda_V+\lambda_W))^\beta}\left(F_{G}(a)-F_{G}(\max\{0, a-t\})\right)\right),
\end{split}
\end{align}
where $F_G$ is the CDF of the Gamma$(\beta, \lambda_V+\lambda_W)$ distribution. By Leibniz integral rule, we may show that , for any given $t>0$, the CDF is an increasing function in $\beta$,
\begin{align*}
\frac{\partial}{\partial\beta}F_{T_{kl}}(t) > 0, \quad{} 0<t<+\infty.
\end{align*}
That is, when two trials have the same enrollment period, the one with larger $\beta$ will have larger probability of observing an event at or before any given time point $t$.

Another observation from equation (\ref{equ:nunif}) is that, if we assume uniform enrollment, namely $\beta=1$, we may simplify 
\begin{align}
\label{equ:unif}
F_{T_{kl}}(t) 
=\frac{\lambda_V}{\lambda_V+\lambda_W}
\left(1 - \max\{0, \frac{a-t}{a}\} - \frac{e^{-(\lambda_V+\lambda_W)t}}{a(\lambda_V+\lambda_W)}\left(e^{(\lambda_V+\lambda_W)(\min(a, t))} - 1\right)\right).
\end{align}
One may verify that, when $t>a$, this formula is the same as Machida et al's formula \cite{machida2021} for exponential survival distribution. Formula (\ref{equ:unif}), however, is more general to include the situation where $t<a$, i.e., when the event of interest occurred  before the enrollment period ends. 

\subsection{Study Duration Comparison between All-comer and Enrichment Designs \label{sect:AvsE}}
\quad{}\newline 
One important application of the theoretical framework is to compare the all-comer and enrichment design in terms of their study duration.

Consider a population of two subgroups $k=1,2$, e.g., biomarker positive vs biomarker negative. Assume a single-arm ($L=1$) study. Let $r_{1}$, $r_{2}$ be the prevalence of the biomarker positive and biomarker negative, respectively. The $\lambda_{1}$, $\lambda_{2}$ are the hazard rates of these two subgroups. Further assume no drop-out and uniform enrollment. The enrollment rates are $m_{1}=r_{1}m$, $m_{2}=r_{2}m$, when $m$ is the enrollment rate of the entire population. Under uniform enrollment, the full study sample size $n=ma$. If we want to enroll the same number of patients in only a subset, e.g., biomarker positive, of the population, the enrollment rate will decrease to $m*r_1$ and therefore, the enrollment period has to extend to $a/r_1$.

For the all-comer design on these two subgroups, the time from study initiation to the observed event is a random variable whose CDF
\begin{align}
\label{equ:TA}
\begin{split}
F_{T_A}(t) &= r_{1}F_{T_{1}}(t) + r_{2}F_{T_{2}}(t)\\
&= 1 - \max\{0, 1-t/a\} - \frac{r_{1}e^{-\lambda_{1}t}}{a\lambda_{1}}\left(e^{\lambda_{1}\min(a, t)} - 1\right) - 
\frac{r_{2}e^{-\lambda_{2}t}}{a\lambda_{2}}\left(e^{\lambda_{2}\min(a, t)} - 1\right).    
\end{split}
\end{align}

For the enrichment design of the same sample size on the biomarker positive patients, its CDF of time from study initiation to observed event is
\begin{align}
\label{equ:TE}
F_{T_E}(t) = 
1 - \max\{0, 1-r_{1}t/a\} - \frac{r_{1}e^{-\lambda_{1}t}}{a\lambda_{1}}\left(e^{\lambda_{1}\min(a/r_{1}, t)} - 1\right),
\end{align}
where the enrollment period is extended to $a/r_{1}>a$ to account for the fact that only a fraction $r_{1}$ of the entire population is eligible for the enrichment trial.

Recall in formula~\ref{equ:Td_est1}, we show that the study duration is approximately $F_T^{-1}(d/n)$. Thus, we may compare the study duration of different designs by examining the CDFs $F_{T_A}$ and $F_{T_E}$. If, for example, we may show that for some $t_E^*>0$,
\begin{align*}
\frac{d}{n} = F_{T_E}(t_E^*) > F_{T_A}(t_E^*),
\end{align*}
then, $t_E^* = F_{T_E}^{-1}(\frac{d}{n})<F_{T_A}^{-1}(\frac{d}{n})=t_A^*$. That is, the enrichment design has shorter duration if its CDF at the time of completion is larger than the CDF of the all-comers design.

Generally, we may derive the following formula for the difference between these two CDFs (Appendix \ref{sect:proof:equ:Tdiff}). 
\begin{align}
\label{equ:Tdiff}
\begin{split}
F_{T_E}(t) - F_{T_A}(t) 
= 
    \begin{cases}
      \frac{1-r_{1}}{a} \left(
      \frac{1 - e^{-\lambda_{2}t}}{\lambda_{2}} - t\right), & 0 < t < a \\
      -(1-r_{1}) + 
      \frac{r_{1}}{a}\left((t-a)-\frac{1-e^{-\lambda_{1}(t-a)}}{\lambda_{1}}\right) + \frac{1-r_{1}}{a}\left(\frac{e^{-\lambda_{2}(t-a)}-e^{-\lambda_{2}t}}{\lambda_{2}}\right), & a < t < a/r_{1} \\
      \frac{r_{1}}{a}\left(\frac{e^{-\lambda_{1}(t-a)}-e^{-\lambda_{1}(t-a/r_{1})}}{\lambda_{1}}\right) + 
      \frac{1-r_{1}}{a}\left(\frac{e^{-\lambda_{2}(t-a)}-e^{-\lambda_{2}t}}{\lambda_{2}}\right), & t > a/r_{1}
    \end{cases}
\end{split}
\end{align}

There are a few remarks may be made regarding the above formula. 

\begin{rmk}
    In the first period, $0<t<a$, the enrollment is ongoing in both trials. Using known inequalities in Appendix \ref{sect:app1}, we may prove that $F_{T_E}(t) - F_{T_A}(t)<0$.. That means, if the all-comer trial completes before the enrollment, i.e., $0<t_A^*<a$, then the enrichment design always has longer duration. Intuitively speaking, if the enrollment is slow such that the target number of events of the all-comer trial can be reached before $a$, we should not expect any enrichment design to be faster. 
\end{rmk}

\begin{rmk}
    In the second period, $a<t<a/r_1$, the all-comer enrollment is completed but the enrichment enrollment is still ongoing. The first term $-(1-r_{1})<0$. The second term $
      \frac{r_{1}}{a}\left((t-a)-\frac{1-e^{-\lambda_{1}(t-a)}}{\lambda_{1}}\right)>0$ is monotone increasing in $r_1$ and $\lambda_1$. The third term $\frac{1-r_{1}}{a}\left(\frac{e^{-\lambda_{2}(t-a)}-e^{-\lambda_{2}t}}{\lambda_{2}}\right)$ is also larger than $0$ but monotone decreasing in $r_1$ and $\lambda_2$. If everything else is equal but the biomarker positive prevalence $r_1$ is higher, then it is possible that $F_{T_E}(t) > F_{T_A}(t)$ for some $t>a$. Thus, the enrichment design has a shorter duration than the all-comers design. Similarly, if $\lambda_1$ is large and $\lambda_2$ is small, i.e., hazard ratio of the biomarker positive, $\frac{\lambda_1}{\lambda_2}$ is large, it is possible for the second term to become larger than the first so that the enrichment design saves time.
\end{rmk}

\begin{rmk}
    In the third period, $t>a/r_1$, both trials have completed the enrollment. Notice that the first term $\frac{r_{1}}{a}\left(\frac{e^{-\lambda_{1}(t-a)}-e^{-\lambda_{1}(t-a/r_{1})}}{\lambda_{1}}\right)$ is less than $0$ and monotone increasing in $r_1$ and $\lambda_1$. Following the same argument in Remark 2, we need a large hazard ratio within the biomarker positive subgroup, $\frac{\lambda_1}{\lambda_2}$, such that the enrichment design is more time-efficient.
\end{rmk}

The derivation for multi-arm studies, $L\geq 2$, and/or more biomarker subgroups, $K>2$, is similar but cumbersome. The comparison between all-comer and the enrichment design will be carried out numerically in the  Section~\ref{sect:comparison}. 

\section{Numerical Results\label{sect:numerical}}
\subsection{The Distributions of $T$ and $T_{(d)}$}\label{sect:T}
\quad{}\newline 
In this section, we will focus on the improvement of the proposed model compared to Machida et al. \cite{machida2021} by demonstrating that formula \ref{equ:nunif} is more general and that there may be situations where this generalness is needed for accurate duration calculation. Here, we define the target number of events $d=88$, sample size $n=140$, no biomarker subgroups, and 1-1 treatment allocation ratio. Two scenarios were considered: 
\begin{enumerate}
\item{Scenario 1:} The enrollment rate is 10 patients per month, or a 14 months enrollment period under uniform enrollment assumption. The survival distributions are assumed to be exponential, and the median survival is 10 and 20 months for the placebo group and the treatment group, respectively.
\item{Scenario 2:} Similar to Scenario 1, however the enrollment rate is modified to be 3.88 patients per month (for a 36 months enrollment period). The median survival changes to 5 and 10 months for the placebo group and the treatment group, respectively. 
\end{enumerate}

\begin{figure}[h]
\centering
\includegraphics[width=8cm, height=6cm]{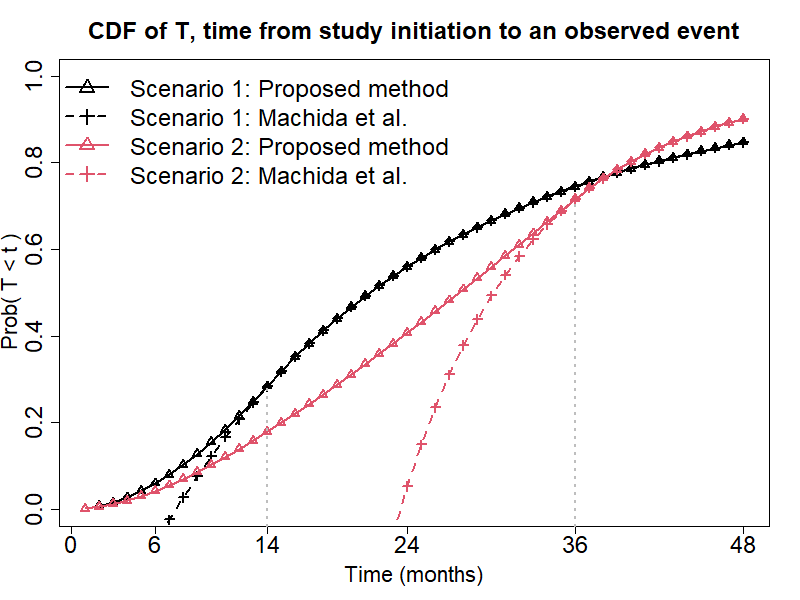}
\includegraphics[width=8cm, height=6cm]{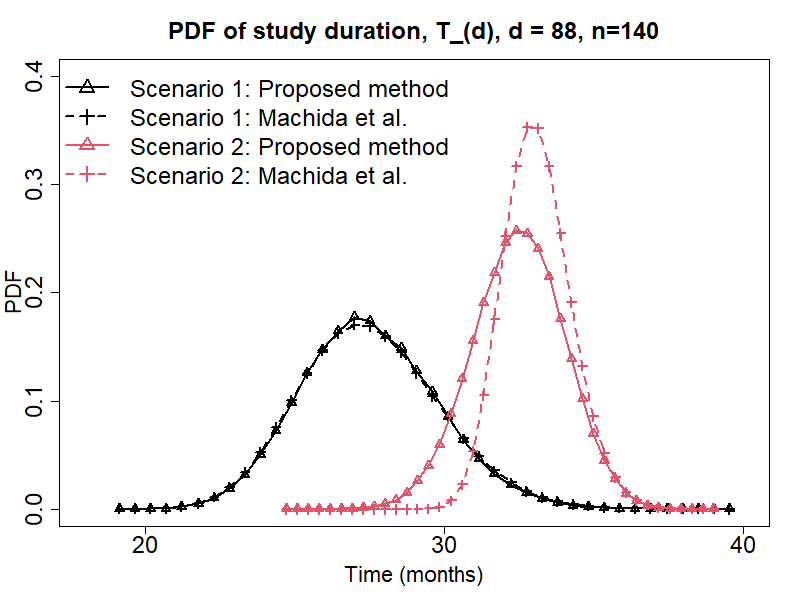}
\caption{\textbf{The distributions of $T$ (left) and $T_{(d)}$ (right)}. Scenario 1: enrollment rate $=10$ pts/month, median survival $=10, 20$ months for the placebo and treatment groups, respectively. Scenario 2: enrollment rate $=3.88$ pts/month, median survival $=5, 10$ months for the placebo and treatment groups, respectively. }
\label{fig1}
\end{figure}

Figure~\ref{fig1} shows that, in terms of the CDF of $T$, the proposed method matches with the Machida et al method when the time of interest is longer than $a$, the enrollment period in both scenarios. However, when $T<a$, the CDF calculated by Machida et al method is no longer a valid CDF. The observation is consistent with the remark made in Machida et al (2021) \cite{machida2021} that they `do not consider the situation in which the study is completed during the enrollment period'.

When comparing these two methods in terms of the distribution of $T_{(d)}$, Figure~\ref{fig1} shows that, in scenario 1, both methods yield almost identical results while, in scenario 2, the two methods differ significantly. The reason is that in scenario 2, there is a non-negligible probability for the study to complete before the enrollment ends due to the slow enrollment and fast disease progression. Therefore, the $(T<a)$ section of the CDF of $T$ is vital for the accurate calculation of the duration $T_{(d)}$.  

\subsection{The Impacts of Non-uniform Enrollment}
\quad{}\newline 
To examine the impact of non-uniform enrollment on study duration, we consider the following variations of the two scenarios in \ref{sect:T}:
\begin{enumerate}
\item{Scenario $1^\prime$:} similar to Scenario 1 but the enrollment follows a Beta(1, 0.45) as in formula~\ref{equ:beta}.
\item{Scenario $2^\prime$:} similar to Scenario 2 but the enrollment follows a Beta(1, 1.25) as in formula~\ref{equ:beta}. 
\end{enumerate}

\begin{figure}[H]
\centering
\includegraphics[width=6cm, height=6cm]{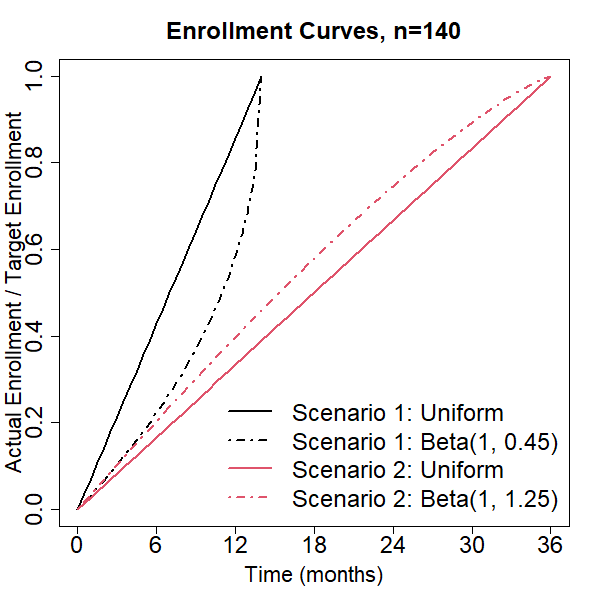}
\includegraphics[width=8cm, height=6cm]{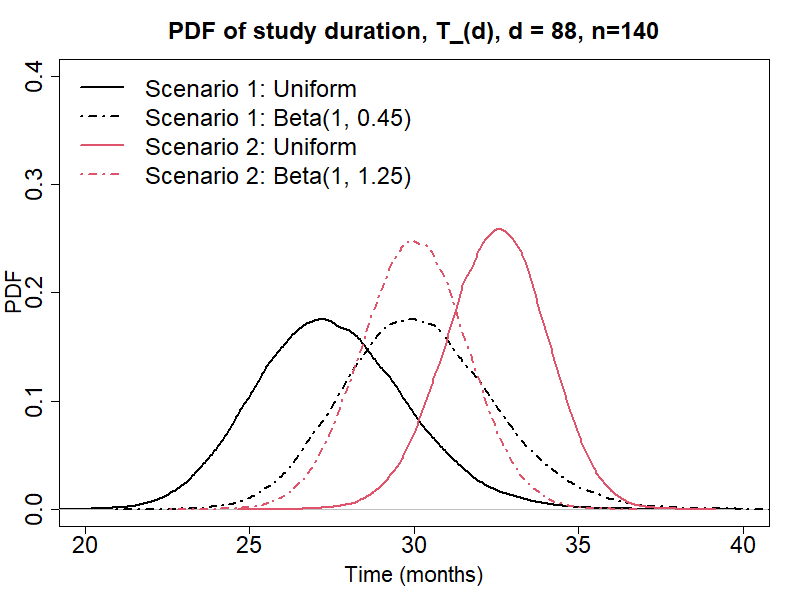}
\caption{\textbf{The enrollment distributions and their impacts on study duration distributions}. Scenario 1: enrollment period = $14$ months, median survival $=10, 20$ months for the placebo, treatment groups, respectively. Scenario 2:  enrollment period = $36$ months, median survival $=5, 10$ months for the placebo, treatment groups, respectively. }
\label{fig2}
\end{figure}

As we discussed in Section~\ref{sect:nuniform} from a theoretical point of view, Figure~\ref{fig2} shows that a Beta(1,$\beta$) distribution with $\beta<1$ means the enrollment starts slower than the average rate. Its enrollment curve (black dotted-dashed) is below the corresponding uniform enrollment curve (black solid), even though these two have the same sample size and enrollment period. On the other hand, if $\beta>1$, the enrollment rate is higher in the beginning than the average. The resulting enrollment curve (red dotted-dashed) is above the corresponding uniform enrollment curve (red solid). 

From the right panel of Figure~\ref{fig2}, we see that, under uniform enrollment, scenario 1 is about 8 months faster than scenario 2. However, under non-uniform enrollment, these two scenarios could have about the same predicted study duration. Specifically, this happens when the enrollment is slower than average ($\beta<1$) in the beginning for scenario 1 and the opposite ($\beta>1$) for scenario 2. 

\subsection{The Impacts of Patient Heterogeneity\label{sect:hetero}}
\quad{}\newline 
In this section, as in Scenario 1, we assume the sample size $n=140$ with 1-1 treatment-placebo allocation, target number of events $d=88$, uniform enrollment and exponential survival distribution. The hazard ratio of the treatment is assumed to be $0.5$, i.e., if the overall median survival time of the placebo group is $MST_{pbo}$, then the overall median survival time in the treatment group $MST_{trt}=2*MST_{pbo}$. 

We introduce patient heterogeneity by assuming the population consists of two groups of patients: the biomarker positive and the biomarker negative. Then, $MST_{trt}$ and $MST_{pbo}$ are from mixtures of biomarker positive and negative patients.  Next, we shall briefly describe how we define the median survival time for each biomarker subgroup.

We may derive, for example, the relationship between $MST_{pbo}$ and  $MST_{pbo, neg}$ (the median survival of biomarker negative patients that received placebo) as the following
\begin{align*}
qe^{-log(2)HR_{pos}\frac{MST_{pbo}}{MST_{pbo, neg}}}+(1-q)e^{-log(2)\frac{MST_{pbo}}{MST_{pbo, neg}}} =  0.5,
\end{align*}
where $q$, $HR_{pos}$ are the prevalence and hazard ratio of the biomarker positive patients, respectively. The biomarker effect is assumed to be prognostic, i.e., $HR_{pos}$ is the same in the treatment and placebo groups. Using the above equation, for given $q$ and $HR_{pos}$, we may solve for $MST_{pbo, neg}$ numerically. Subsequently, we may calculate $MST_{pbo, pos}=MST_{pbo, neg}/HR_{pos}$. Then,  $MST_{trt, neg}$ and $MST_{trt, pos}$ may be obtained in a similar fashion. To model the extent of heterogeneity, we set the prevalence $q$ ranges from $10\%$ to $90\%$ and the hazard ratio $HR_{pos}=1,2,...,5$. 

\begin{figure}[h]
\centering
\includegraphics[width=5.3cm, height=5cm]{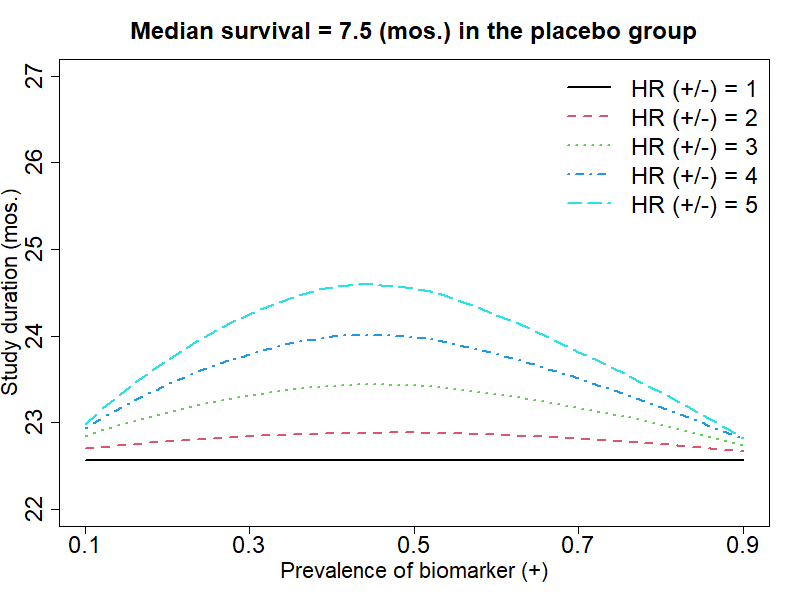}
\includegraphics[width=5.3cm, height=5cm]{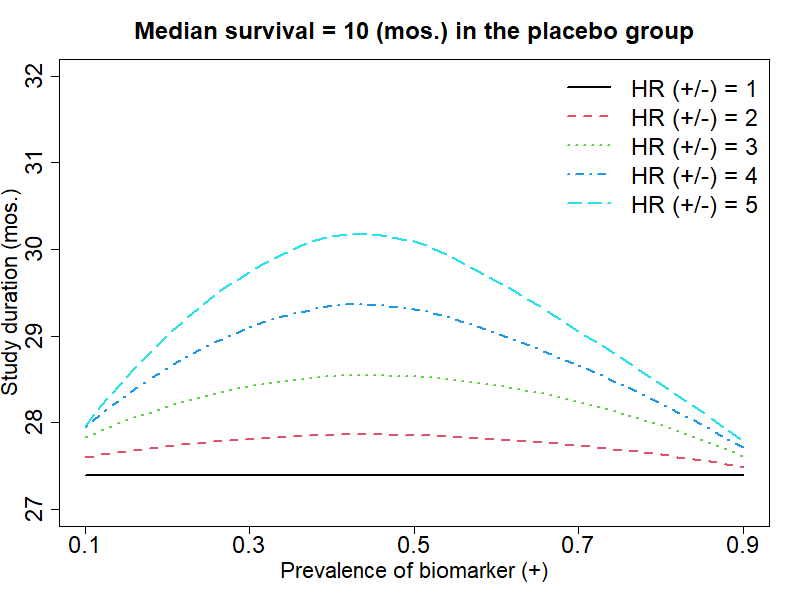}
\includegraphics[width=5.3cm, height=5cm]{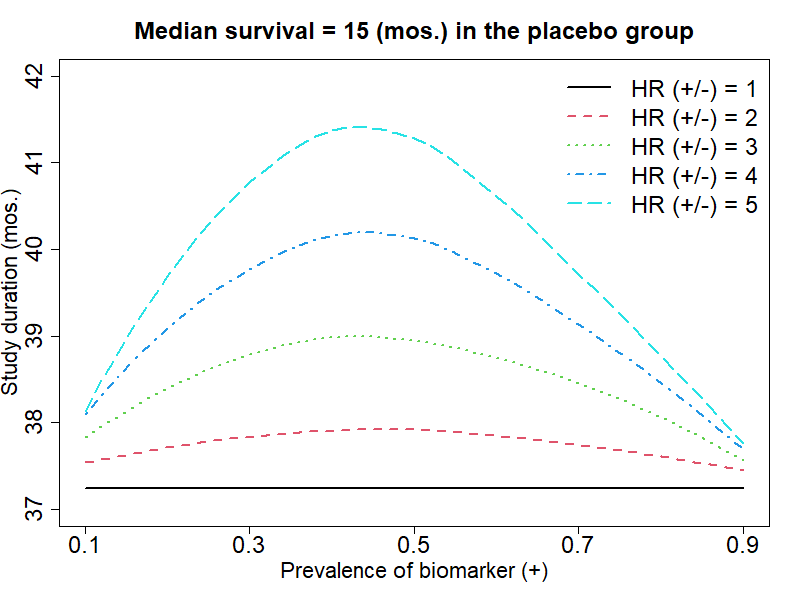}
\caption{\textbf{The impacts of patient heterogeneity on study duration}. The enrollment is assumed to be uniform. Left: median survival $MST_{pbo}=7.5$ months, enrollment rate $=7$ patients/month. Middle: median survival $MST_{pbo}=10$ months, enrollment rate $=10$ patients/month. Right: median survival $MST_{pbo}=15$ months, enrollment rate $=20$ patients/month.}
\label{fig3a}
\end{figure}

The study duration for each pair of the prevalence and hazard ratio of the biomarker is summarized in Figure~\ref{fig3a}. The black solid line is the benchmark where the hazard ratio is $1$, i.e., no biomarker effect. Thus, the line is perfectly horizontal regardless of the prevalence. When the biomarker hazard ratio increases, the colored curves begin to show a concaved down behavior, showing that the study duration increases when the biomarker effect is present even if the median survival times are the same for every point along each curve. Ignoring the biomarker subgroups will result in underestimating the study duration of the traditional placebo-controlled, all-comers design by up to 10\% across different median survival times and enrollment rates that we examined. The maximum discrepancy occurred when the prevalence is about $45\%$ and the hazard ratio, as expected, is the largest at $5$. 

\subsection{Comparison between the All-comers Design and the Enrichment Design \label{sect:comparison}}
\begin{figure}[h]
\centering
\includegraphics[width=5.3cm, height=4cm]{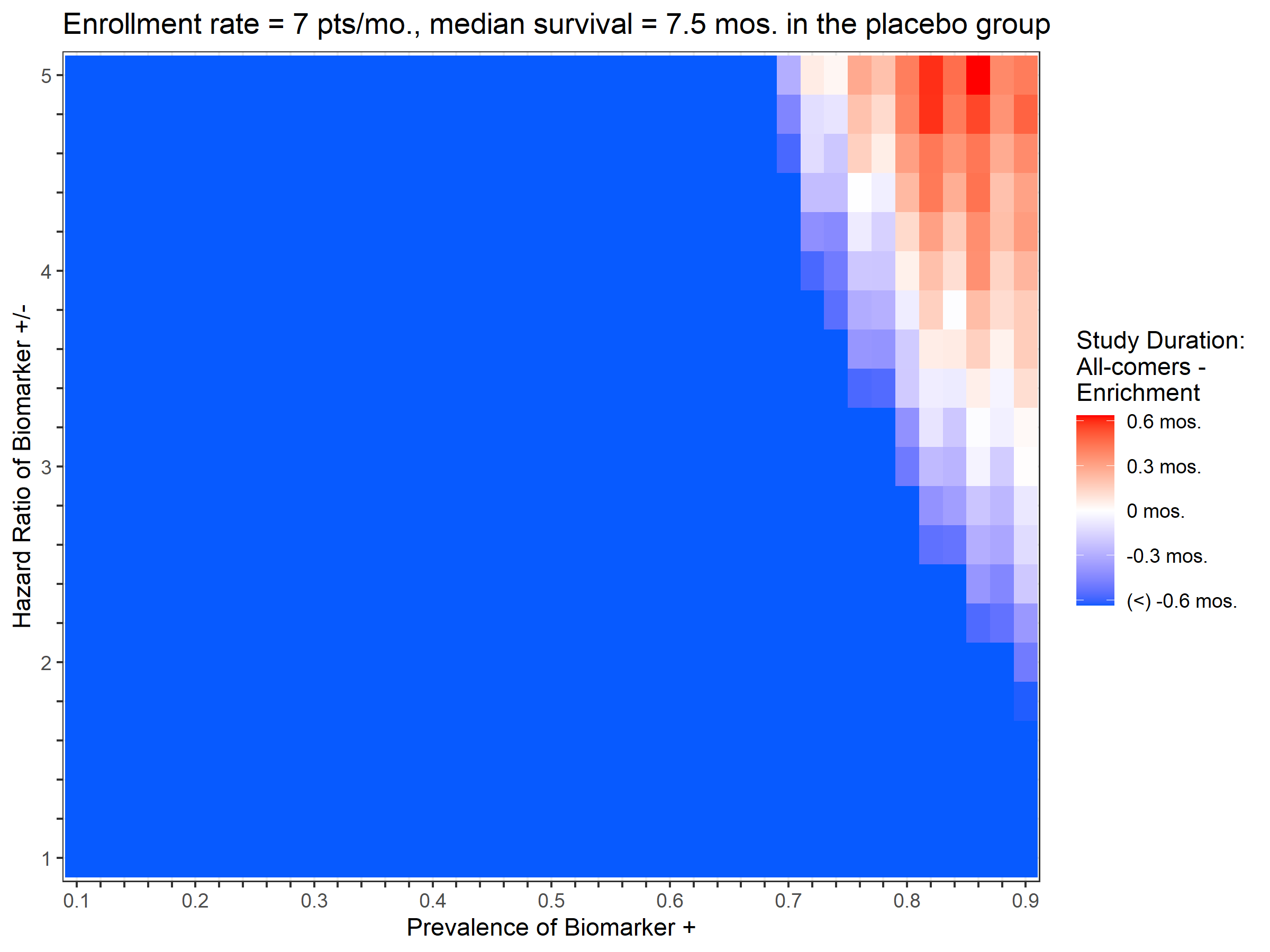}
\includegraphics[width=5.3cm, height=4cm]{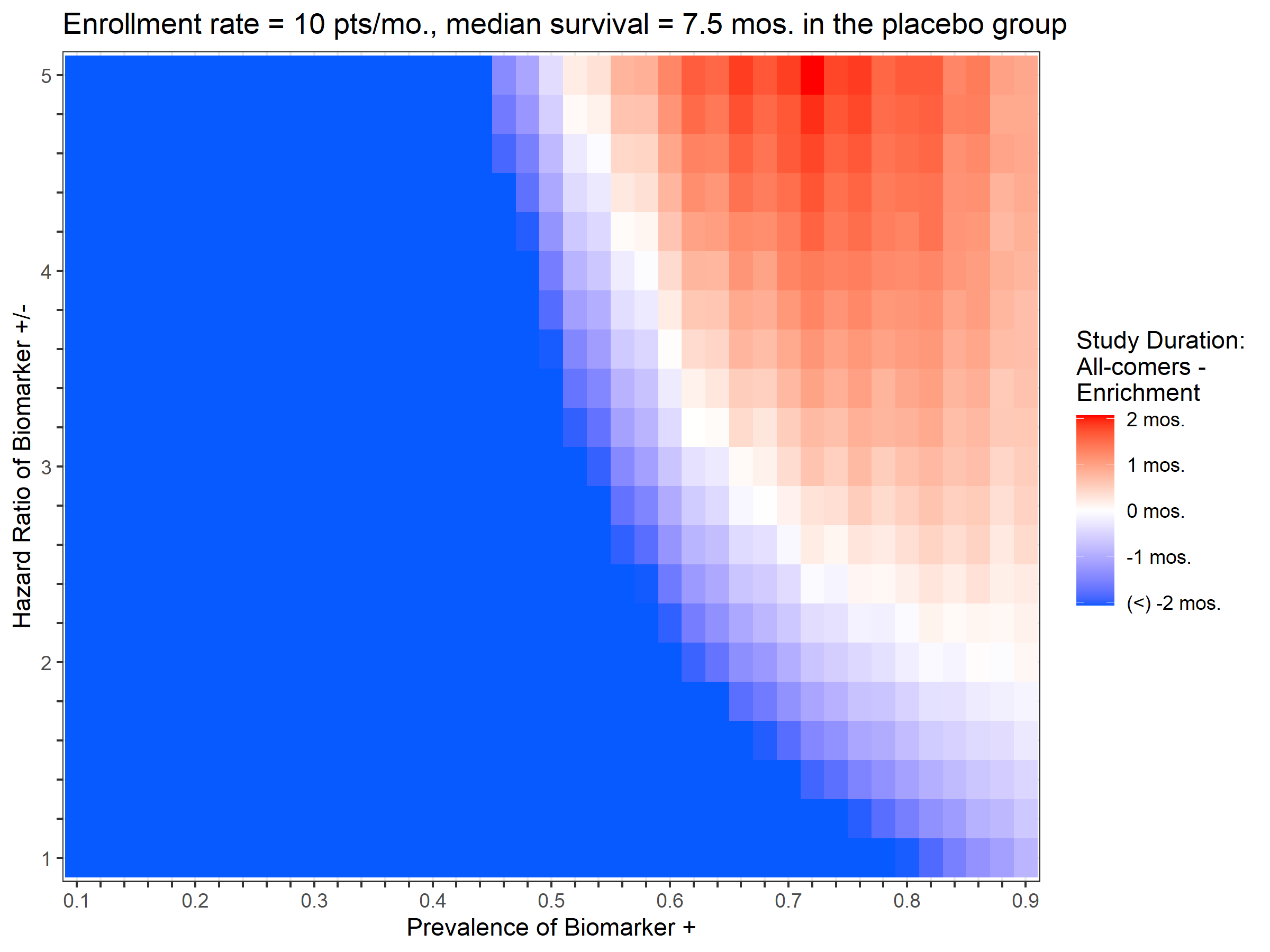}
\includegraphics[width=5.3cm, height=4cm]{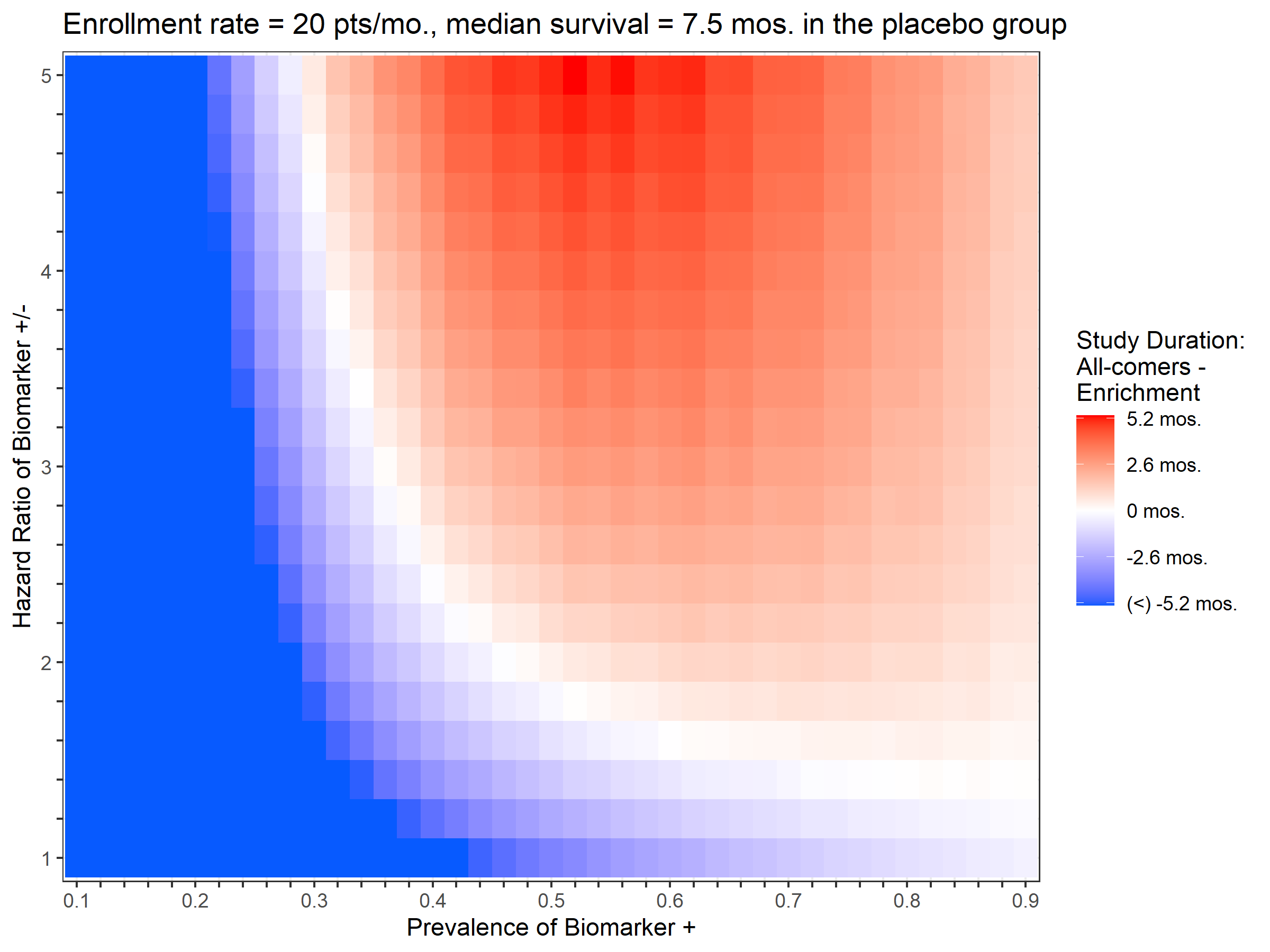}\\
\includegraphics[width=5.3cm, height=4cm]{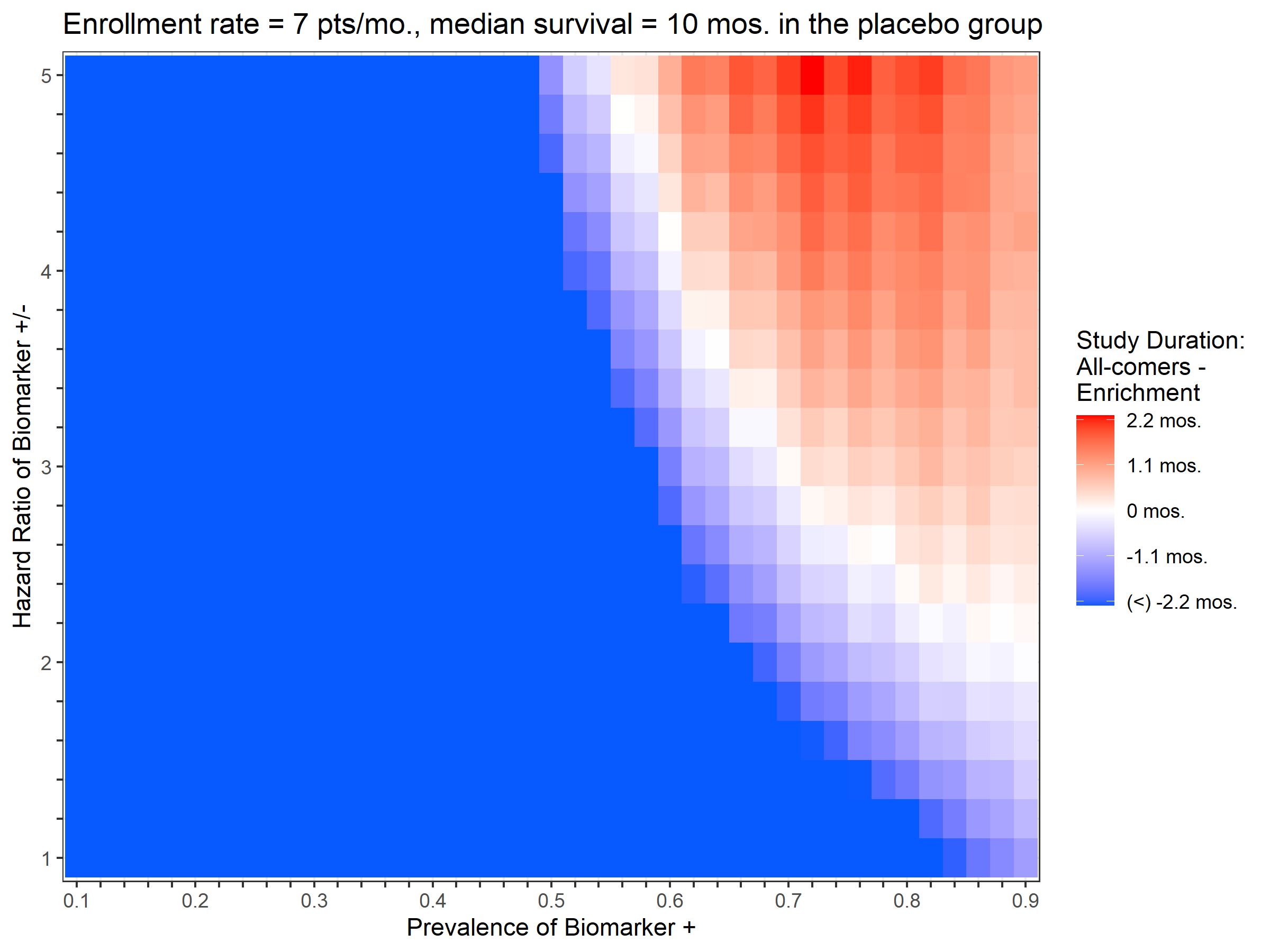}
\includegraphics[width=5.3cm, height=4cm]{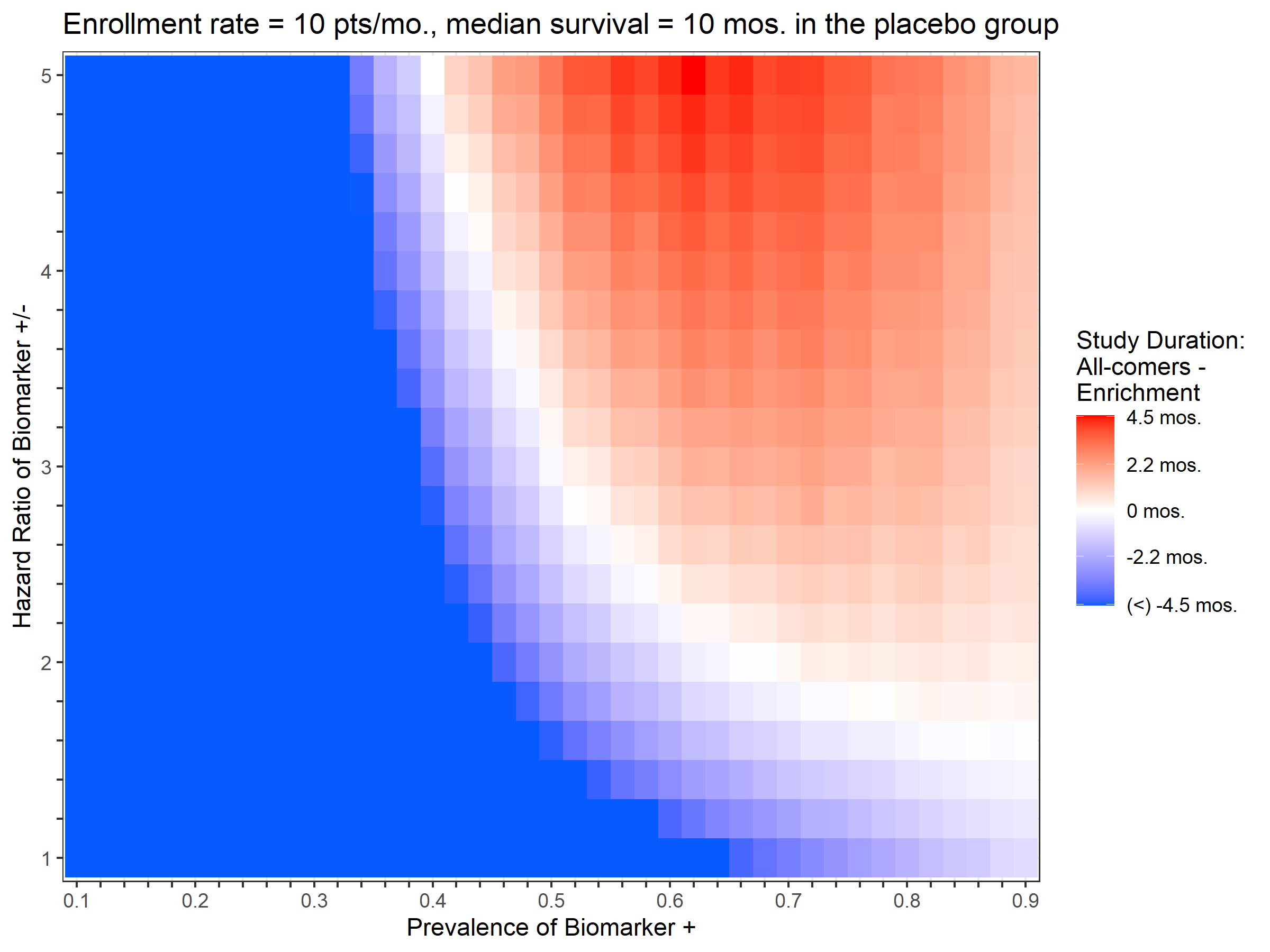}
\includegraphics[width=5.3cm, height=4cm]{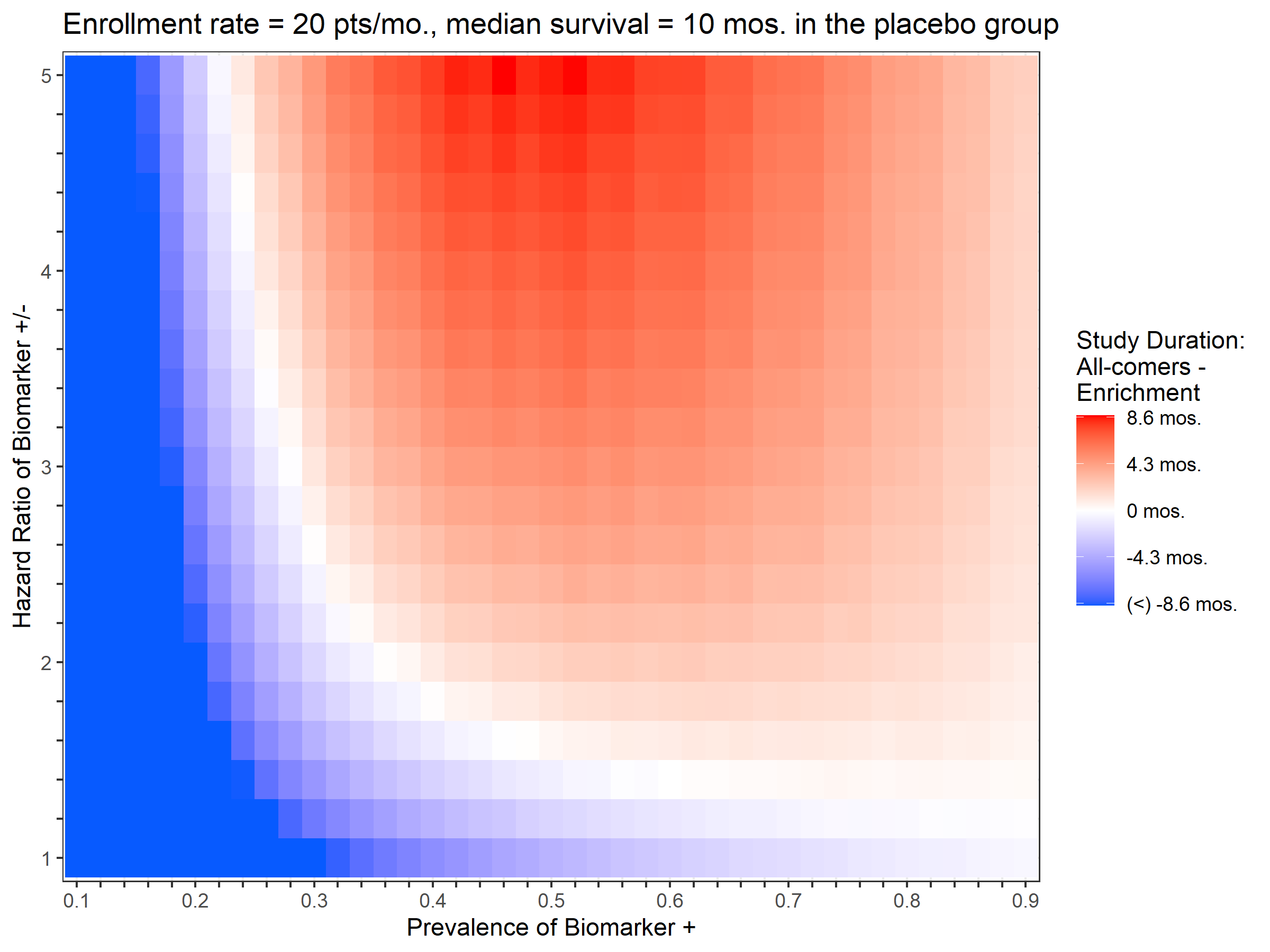}\\
\includegraphics[width=5.3cm, height=4cm]{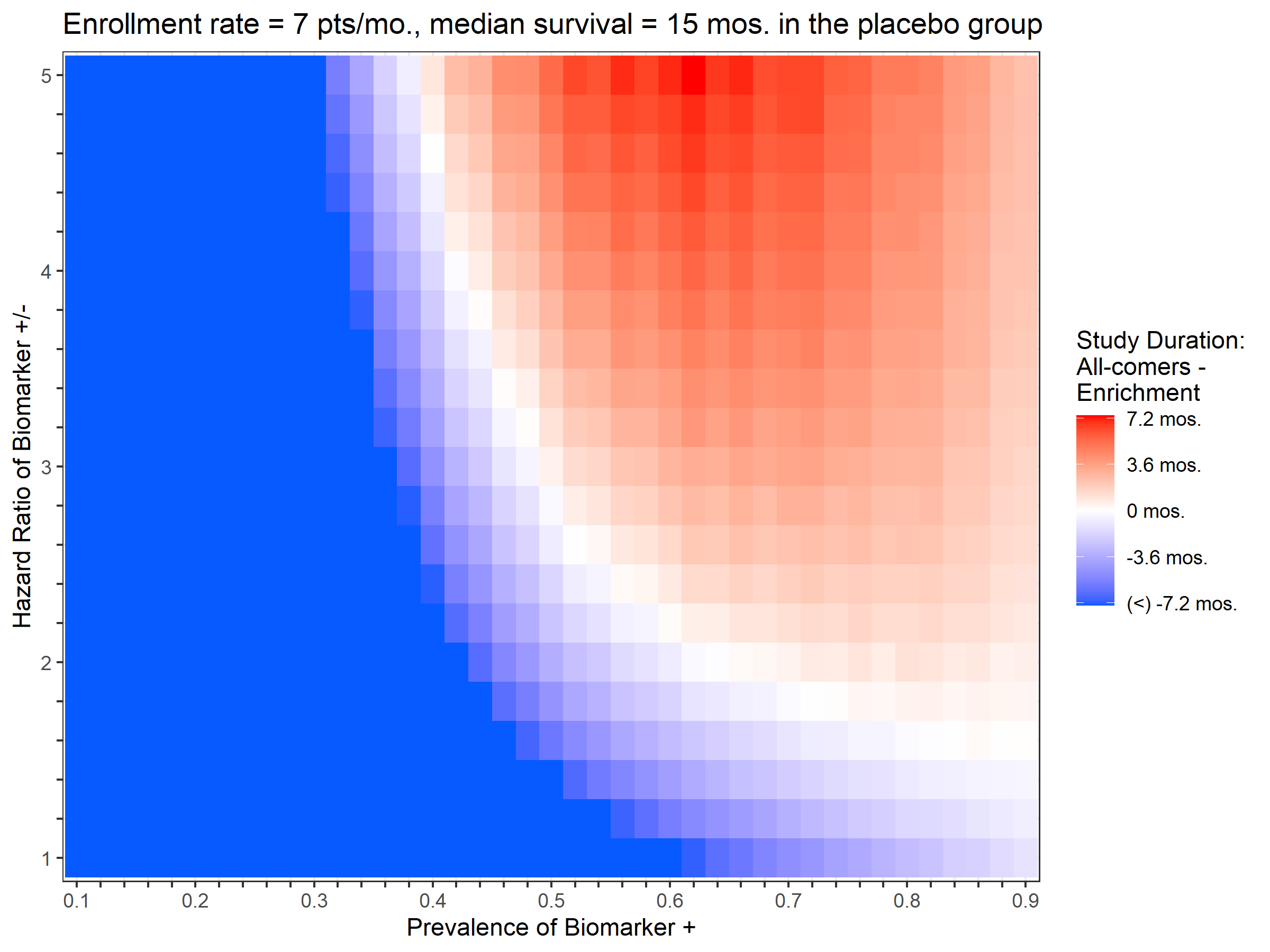}
\includegraphics[width=5.3cm, height=4cm]{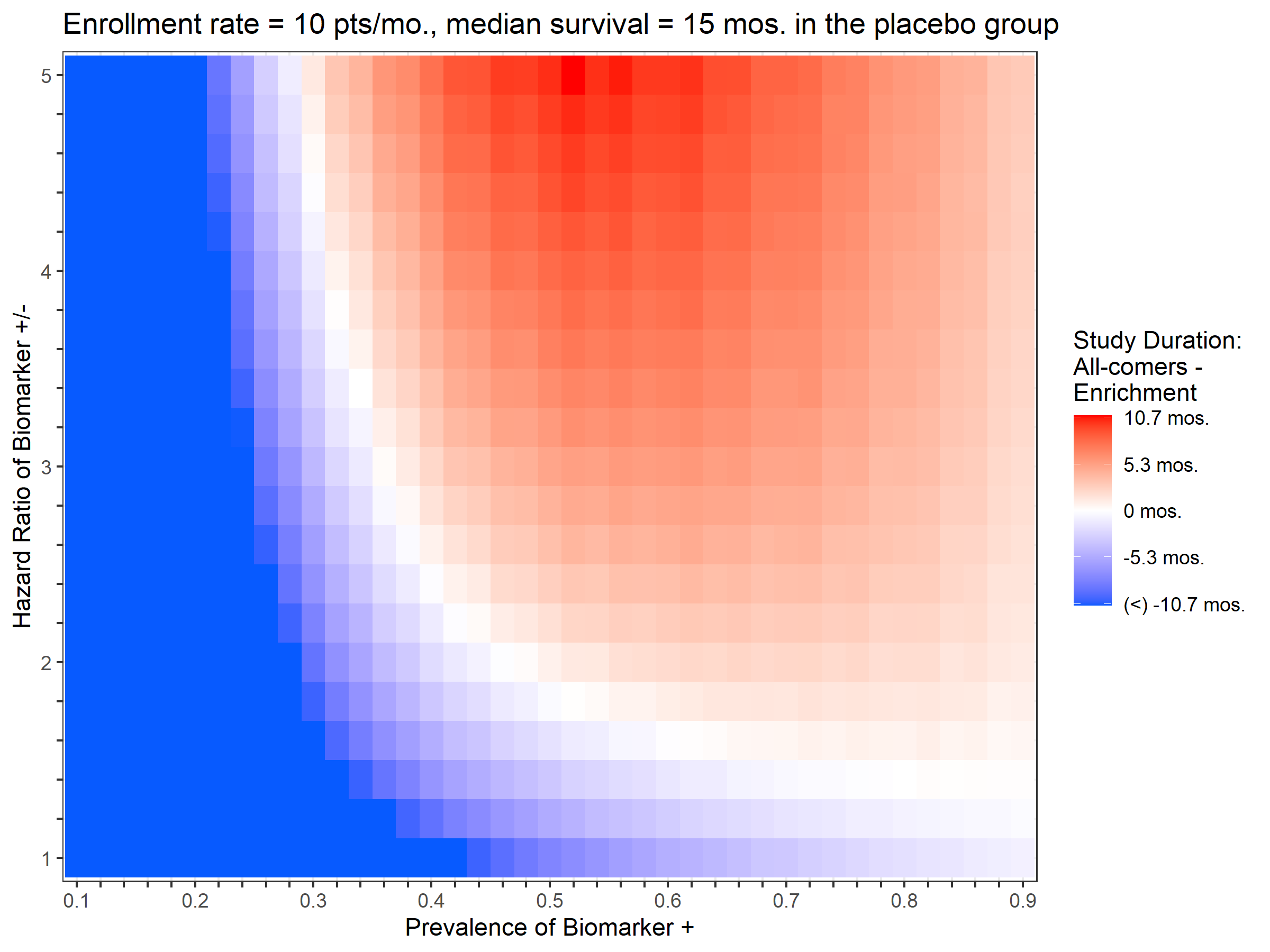}
\includegraphics[width=5.3cm, height=4cm]{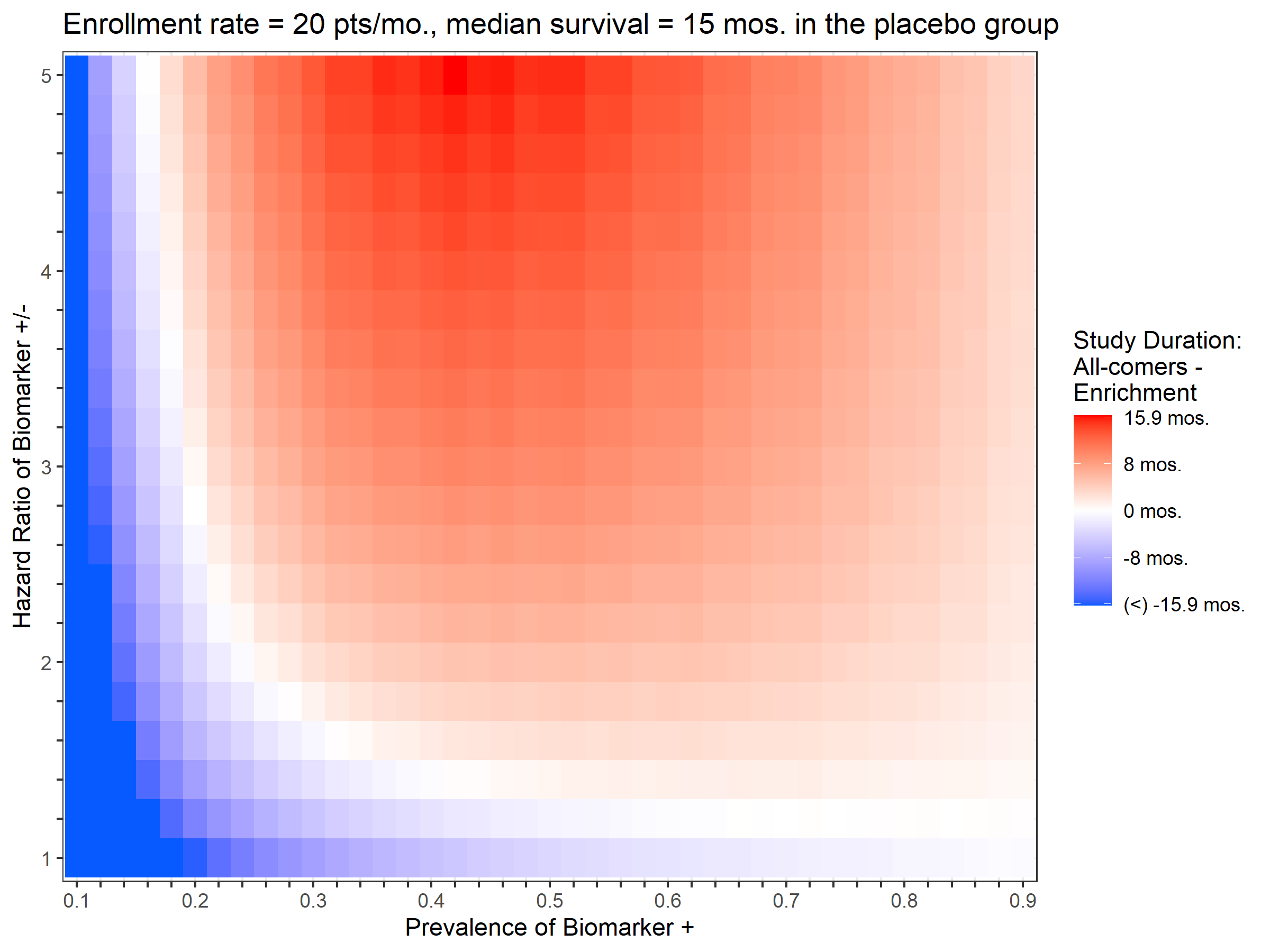}
\caption{The heatmap of the study duration difference between the all-comers design and the enrichment design.  Columns: the enrollment rate $=7, 10, 20$ patients per month. Rows: median survival time of the placebo group $=7.5, 10, 15$ months. within each panel, x-axis: the prevalence of the biomarker positive; y-axis: hazard ratio of the biomarker positive.}
\label{fig4}
\end{figure}

\quad{}\newline 
One of the important applications of the proposed calculation method is to compare the study duration of the all-comers design and the biomarker enrichment design. Here, we keep the settings of the all-comers' design the same as in Section~\ref{sect:hetero}. For the enrichment design, the difference is that only biomarker positive patients are enrolled over a longer period, $a/q>a$, where $a$ is the enrollment period of the all-comers design and $q$ is the prevalence of the biomarker positive. For each pair of prevalence and hazard ratio of the biomarker, we calculate the difference of the study duration of the all-comers design and the enrichment design. The results are summarized in the heatmaps in Figure~\ref{fig4}. 

There are multiple interesting observations. First, there seems to exist a decreasing function in the prevalence - hazard ratio space as the boundary that separates the region (in red) where the enrichment is more time-efficient and the opposite region (in blue). Second, the enrichment-favoring region is larger when the enrollment rate is quicker and/or the median survival is longer. For example, when the median survival is $15$ months and $20$ patients may be enrolled per month, then an enrichment design based on a mild biomarker effect, such as a hazard ratio of $2$ with $30\%$ prevalence, already has a shorter duration than the all-comers design. Lastly, the white boundary seems to have steep slope from the beginning, which means a decrease in the prevalence has to be compensated by a larger increase in the hazard ratio. Therefore, the prevalence seems to be a more important factor than the hazard ratio when we are choosing the biomarker for enrichment. 

Additionally, we also considered the comparison in the space of enrollment rate and median survival time. Figure~\ref{fig:supp1} summarized the results in a similar format, where we varied the enrollment rate from $7$ to $20$ patients per month and the median survival time from $7.5$ to $15$ months in a finer grid of values. The observations are similar: 1) There exists a boundary in terms of the enrollment rate and median survival, where the enrichment is favorable in the region above the boundary. For example, if the enrollment is fast and median survival is long, then the enrichment design can save a significant amount of time. 2) A biomarker with higher prevalence and stronger effect makes the enrichment-favorable region larger.

\section{Real Data Analysis\label{sect:realdata}}
To illustrate the accuracy and practical utility of the proposed study duration calculation method, we applied it to re-assess two publicly available clinical trial datasets. In a hypothetical randomized, double-blind, placebo-controlled trial setting, we consider two designs: one is a biomarker enrichment design where patients are included only if they belong to a certain biomarker positive group, while the traditional all-comers design does not impose this inclusion criteria.  Next, for a given sample size $n$ and, a given number, $d$, of events, we briefly describe the procedures of the re-assessment. 
\begin{enumerate}
\item To extract the \emph{actual} study duration, we choose, from the real clinical data, the first $n$ patients in the biomarker positive population (or all-comers population) as our hypothetical trial dataset in the enrichment (all-comers) design. Then, we record the $d$th event date of the hypothetical trial dataset. The difference between the first patient enrollment date and the $d$th event date is the trial duration.
\item To \emph{calculate} the study duration, we estimate the survival curves of each subgroups (biomarker-by-treatment) using the hypothetical trial dataset by assuming a Weibull distribution. The enrollment distribution is estimated using the same data by assuming a beta distribution in formula \ref{equ:beta}.
\end{enumerate}

The purpose of this application is to show that under reasonable distributional assumptions and the availability of representative historical data, the proposed model may indeed calculate the study duration accurately.

\subsection{Example 1: \emph{adtte} Data from the R \emph{visR} Package \cite{visR-package}}
\quad{}\newline 
The \emph{adtte} data is a legacy data used in the CDISC SDTM/ADaM Pilot Project \cite{cdisc} that was provided by Eli Lilly and Company from a phase II clinical trial of the safety and efficacy of the Xanomeline Transdermal Therapeutic System (TTS) in patients with mild to moderate Alzheimer’s disease. 

For the purpose of this paper, we are interested in the time to first dermatologic adverse event. We selected the lower Xanomeline dose for the treatment group, resulting in a dataset of 170 patients (88 in the treatment group, 82 in the placebo group). Although not statistically significant, the hazard ratio for male patients compared to females is 1.39. For illustrative purposes, we selected male as the biomarker for our enrichment design, taking into account the prevalence of $39.4\%$. The hypothetical trial dataset will have a sample size of 67, which is the number of male patients in the \emph{adtte} after filtering the high dose treatment.

The results are summarized in the left panel of Figure~\ref{fig:real_data}. The black solid curve is universally higher than the red solid curve, meaning for any given number of events, the all-comer design will always have shorter duration than the enrichment design. On the other hand, given any desired duration, the all-comer design is always able to achieve a higher number of events, which could translate to higher power. The biomarker enrichment result should not be surprising since the biomarker effect (HR=1.39) is not strong enough to compensate the difficulty of recruiting biomarker positive (male only) patients. Another important observation is that the calculated duration (dashed) from the proposed model is close to the actual duration, which shows the accuracy of the calculation algorithm.

\subsection{Example 2: \emph{udca} Data from the R \emph{survival} Package \cite{survival-package}}
\quad{}\newline 
The \emph{udca} data came from a double-blind, placebo-controlled, randomized trial of ursodeoxycholic acid (UDCA) in patients with primary biliary cirrohosis \cite{lindor1994udca, survival-book}. The primary endpoint is the time to treatment failure, which is defined as death, liver transplantation, histological progression or other conditions listed in \cite{lindor1994udca}. The baseline bilirubin level was found to be strongly associated with the time to treatment failure. In this example, we define the biomarker positive patients as those with baseline bilirubin > 1 mg/dL. The resulting biomarker positive subgroup has a hazard ratio of $2.59$ and prevalence $50\%$. Similar to Example 1, we limit the sample size to 84, which is the total number of biomarker positive patients that are available in \emph{udca}.

\begin{figure}[H]
\centering
\includegraphics[width=8cm, height=6cm]{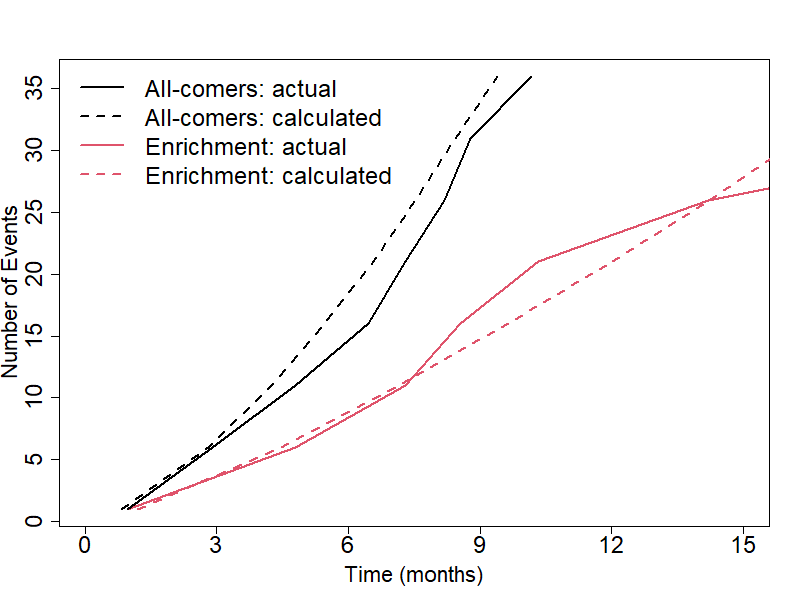}
\includegraphics[width=8cm, height=6cm]{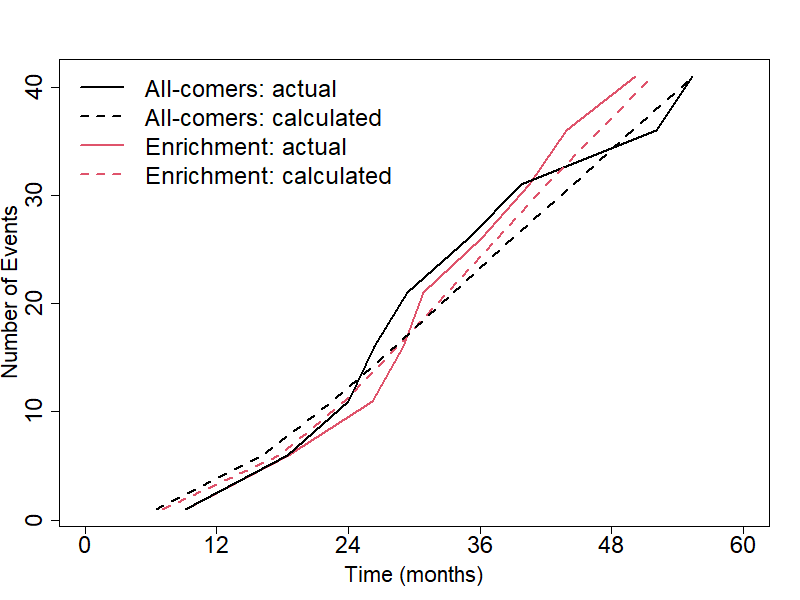}
\caption{The study duration of the hypothetical all-comers and enrichment designs. Left: \emph{adtte} data from R \emph{visR} package, sample size $n=67$. Right: \emph{udca} data from R \emph{survival} package, sample size $n=84$.}
\label{fig:real_data}
\end{figure}

From the right panel of Figure~\ref{fig:real_data}, we may see that the all-comers design has a relatively shorter duration when the required number of events is low, e.g., $d<30$. If the trial targets more than 30 events, however, the enrichment will save as much as 10 months. The calculated durations reflect the transition, starting from $d\approx 20$, of the enrichment design being more time-efficient as the required number of events increases. The overlapping calculated curves and actual curves also show the potential utility of the proposed algorithm to inform study planning.

\section{Discussion\label{sect:discussion}}

The proposed model is a generalization of the order statistic model in \cite{machida2021}. We provided a theoretically rigorous foundation for the derivation of the CDF of study duration $T_{(d)}$. Our model is more general by considering the patient heterogeneity induced by the biomarker effect. We also proposed an algorithm based on simulation that covers a a wide variety of survival distribution and enrollment assumptions. The empirical results show a more robust performance compared to the reference. Several interesting examples and real datasets are also examined to show the broad application potential of the proposed method, such as the impact of non-uniform enrollment on the study duration and the comparison between the enrichment design and traditional all-comers design. 

Although we mainly focused on calculating study duration, the proposed algorithm may be used to answer different questions, such as, given a target duration, how many samples do we need; what is the minimum enrollment rate; or what are the desirable  biomarker characteristics to achieve a certain amount of duration reduction by adopting an enrichment design? This is because our proposed model connects the study duration with other parameters of the disease, trial and patient population. Numerically, the proposed algorithm may be used to search for the parameter values, e.g., via bisection search, that achieve desirable study duration conditions.

One of the major motivations of this research is to quantify the benefits of biomarker guided designs to improve trial efficiency. By integrating innovative biomarker-guided strategies, clinical trials can become more efficient, reducing costs and timelines, and ultimately accelerating the availability of new treatments for patients in need. As mentioned in the Introduction, many such innovative designs have a component of biomarker enrichment. By accurately predicting the (dis)advantages of this enrichment component relative to the traditional all-comers component, we are able to select the most appropriate design under reasonable assumptions and historical data.

An example of the innovative biomarker strategies is the utilization of circulating tumor DNA (ctDNA), which holds significant promise in expediting clinical trials across various cancer types. CtDNA, originating from tumor cells, provides a non-invasive means to monitor the molecular changes within a patient's tumor. Pre-treatment ctDNA variant allele frequency (VAF) has been found to be a prognostic biomarker across many cancer types \cite{pietrasz2017plasma, wang2019prognostic}. A higher VAF often suggests higher tumor burden, therefore a higher chance of progression or death. The enrichment of patients with high baseline ctDNA may allow for faster readout of trials and make the efficacious treatment options available quicker. As the field of ctDNA analysis continues to advance, its potential to guide clinical trial design is becoming increasingly evident. Our proposed method will help with this process by providing statistically credible and quantifiable evidence to support innovative trial designs. 

Lastly, although the proposed model may accommodate both prognostic and predictive biomarkers, most of our discussions focus on prognostic ones where the treatment effect are similar between the patient subgroups. Many innovative biomarker-guided trial designs, however, focused on predictive biomarker where a differential treatment effect is assumed for the biomarker positive and negative groups. In reality, a biomarker may be both prognostic and predictive. For example, high risk patients defined by a prognostic marker may have better or worse response to treatment than low risk patients. A future research direction is to explore the impact of the additional parameters of the effect of predictive biomarkers on the study duration comparison under the proposed framework.

\section{Acknowledgements} 

The authors would like to thank Jean-Francois Martini and Douglas Laird for helpful discussions.

% \section{Bibliography}
\printbibliography

\section*{Appendix}
\renewcommand{\thesubsection}{A\arabic{subsection}}
\renewcommand{\thefigure}{A\arabic{figure}}
\setcounter{figure}{0}

\subsection{Derivation of Equation~(\ref{equ:FTkl})\label{sect:proof:equ:FTkl}}
\begin{align*}
F_{T_{kl}}(t) = P(T_{kl}\leq t) &= P(T_{kl}\leq t, V_{kl}\leq W_{kl}) + P(T_{kl}\leq t, V_{kl}>W_{kl})\\ 
&= P(U_{kl} + V_{kl}\leq t, V_{kl}\leq W_{kl}) + P(+\infty\leq t, V_{kl}>W_{kl})\\
&= P(U_{kl} + V_{kl}\leq t, V_{kl}\leq W_{kl})\\
&=\int_{0}^{\min\{t, a\}}P(V_{kl}\leq t - u, V_{kl}\leq W_{kl})f_{U_{kl}}(u)du\\
&=\int_0^{\min\{t, a\}}\int_{0}^{t-u}\int_{v}^{+\infty} f_{W_{kl}}(w)f_{V_{kl}}(v)f_{U_{kl}}(u)dwdvdu
\end{align*}

\subsection{Derivation of Equation~(\ref{equ:nunif})\label{sect:proof:equ:nunif}}
\begin{align*}
F_{T_{kl}}(t) = & \int_0^{\min\{t, a\}}\int_{0}^{t-u}\int_{v}^{+\infty} f_{W_{kl}}(w)f_{V_{kl}}(v)f_{U_{kl}}(u)dwdvdu \\
= &\int_0^{\min\{t, a\}}\left(\int_{0}^{t-u} e^{-\lambda_Wv} \lambda_Ve^{-\lambda_Vv} dv\right)\frac{\beta}{a} \left(1-\frac{u}{a}\right)^{\beta-1}du\\     
= &\int_0^{\min\{t, a\}}\frac{\lambda_V}{\lambda_V+\lambda_W}\left(1-e^{-(\lambda_V+\lambda_W)(t-u)}\right)\frac{\beta}{a} \left(1-\frac{u}{a}\right)^{\beta-1}du\\  
= &\frac{\lambda_V}{a^\beta(\lambda_V+\lambda_W)}
\int_{\max\{0, a-t\}}^a\left(1-e^{-(\lambda_V+\lambda_W)(x+t-a)}\right) \beta x^{\beta-1}dx\\  
= &\frac{\lambda_V}{\lambda_V+\lambda_W}
\left(1 - \max\{0, 1-t/a\}^\beta - e^{-(\lambda_V+\lambda_W)(t-a)}\frac{\beta\Gamma(\beta)}{(a(\lambda_V+\lambda_W))^\beta}\left(F_{G}(a)-F_{G}(\max(0, a-t))\right)\right)\\  
\end{align*}

\subsection{Derivation of Equation (\ref{equ:Tdiff})\label{sect:proof:equ:Tdiff}}
\begin{align*}
&F_{T_E}(t) - F_{T_A}(t) \\
=& \max\{0, 1-\frac{t}{a}\} - 
\max\{0, 1-r_{11}\frac{t}{a}\} + \frac{r_{11}e^{-\lambda_{11}t}}{a\lambda_{11}}(e^{\lambda_{11}\min(a, t)} - e^{\lambda_{11}\min(\frac{a}{r_{11}}, t)}) + 
\frac{r_{21}e^{-\lambda_{21}t}}{a\lambda_{21}}\left(e^{\lambda_{21}\min(a, t)} - 1\right)  
\\
= & 
    \begin{cases}
      -(1-r_{11})\frac{t}{a} +
      0 + 
      \frac{r_{21}e^{-\lambda_{21}t}}{a\lambda_{21}}\left(e^{\lambda_{21}t} - 1\right), & 0 < t < a \\
      -(1-r_{11}\frac{t}{a}) + 
      \frac{r_{11}e^{-\lambda_{11}t}}{a\lambda_{11}}\left(e^{\lambda_{11}a}-e^{\lambda_{11}t}\right) + \frac{r_{21}e^{-\lambda_{21}t}}{a\lambda_{21}}\left(e^{\lambda_{21}a} - 1\right), & a < t < a/r_{11}\\
      0 + 
      \frac{r_{11}e^{-\lambda_{11}t}}{a\lambda_{11}}\left(e^{\lambda_{11}a}-e^{\lambda_{11}a/r_{11}}\right) + 
      \frac{r_{21}e^{-\lambda_{21}t}}{a\lambda_{21}}\left(e^{\lambda_{21}a} - 1\right), & t > a/r_{11}
    \end{cases}\\
= & 
    \begin{cases}
      \frac{1-r_{11}}{a} \left(
      \frac{1 - e^{-\lambda_{21}t}}{\lambda_{21}} - t\right), & 0 < t < a \\
      -(1-r_{11}) - 
      \frac{r_{11}}{a}\left(\frac{1-e^{-\lambda_{11}(t-a)}}{\lambda_{11}}-(t-a)\right) + \frac{1-r_{11}}{a}\left(\frac{e^{-\lambda_{21}(t-a)}-e^{-\lambda_{21}t}}{\lambda_{21}}\right), & a < t < a/r_{11} \\
      \frac{r_{11}}{a}\left(\frac{e^{-\lambda_{11}(t-a)}-e^{-\lambda_{11}(t-a/r_{11})}}{\lambda_{11}}\right) + 
      \frac{1-r_{11}}{a}\left(\frac{e^{-\lambda_{21}(t-a)}-e^{-\lambda_{21}t}}{\lambda_{21}}\right), & t > a/r_{11}
    \end{cases}
\end{align*}

\subsection{Some Inequalities\label{sect:app1}}
\begin{enumerate}
    \item If $b>0$, $x\geq0$, then $\frac{1-e^{-bx}}{b}-x<0$.
    \item Define $f(x)=\frac{e^{-bx}-e^{-cx}}{x}$, $x\geq 0$.\\
    If $b>c>0$, then $f(x)$ is a monotone increasing function, $c-b<f(x)<0$. \\  
    If $c>b>0$, then $f(x)$ is a monotone decreasing function, $0<f(x)<c-b$.       
\end{enumerate}

\subsection{Supplement Figures}
\quad{}
\begin{figure}[!htb]
\centering
\includegraphics[width=5.3cm, height=4cm]{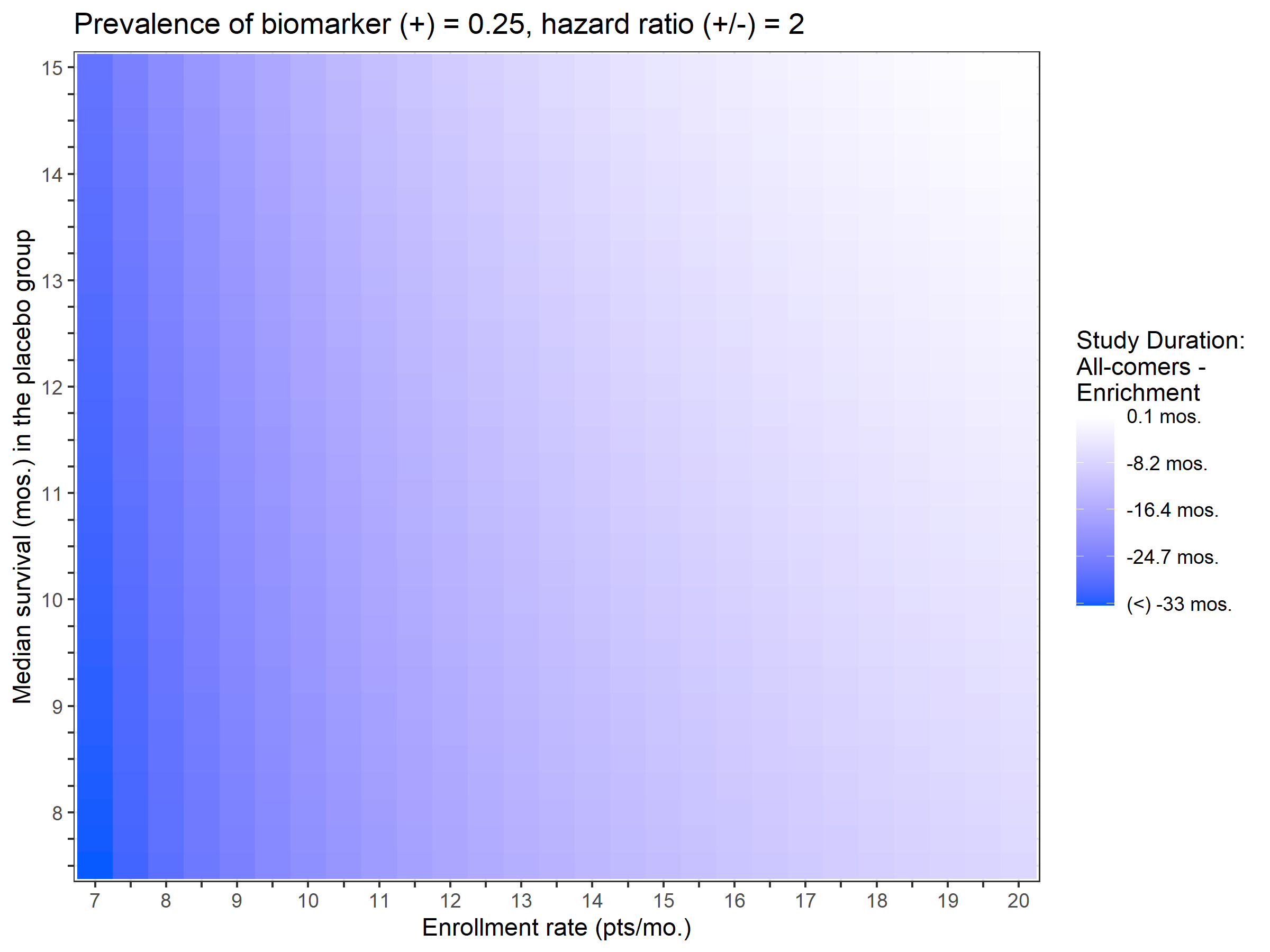}
\includegraphics[width=5.3cm, height=4cm]{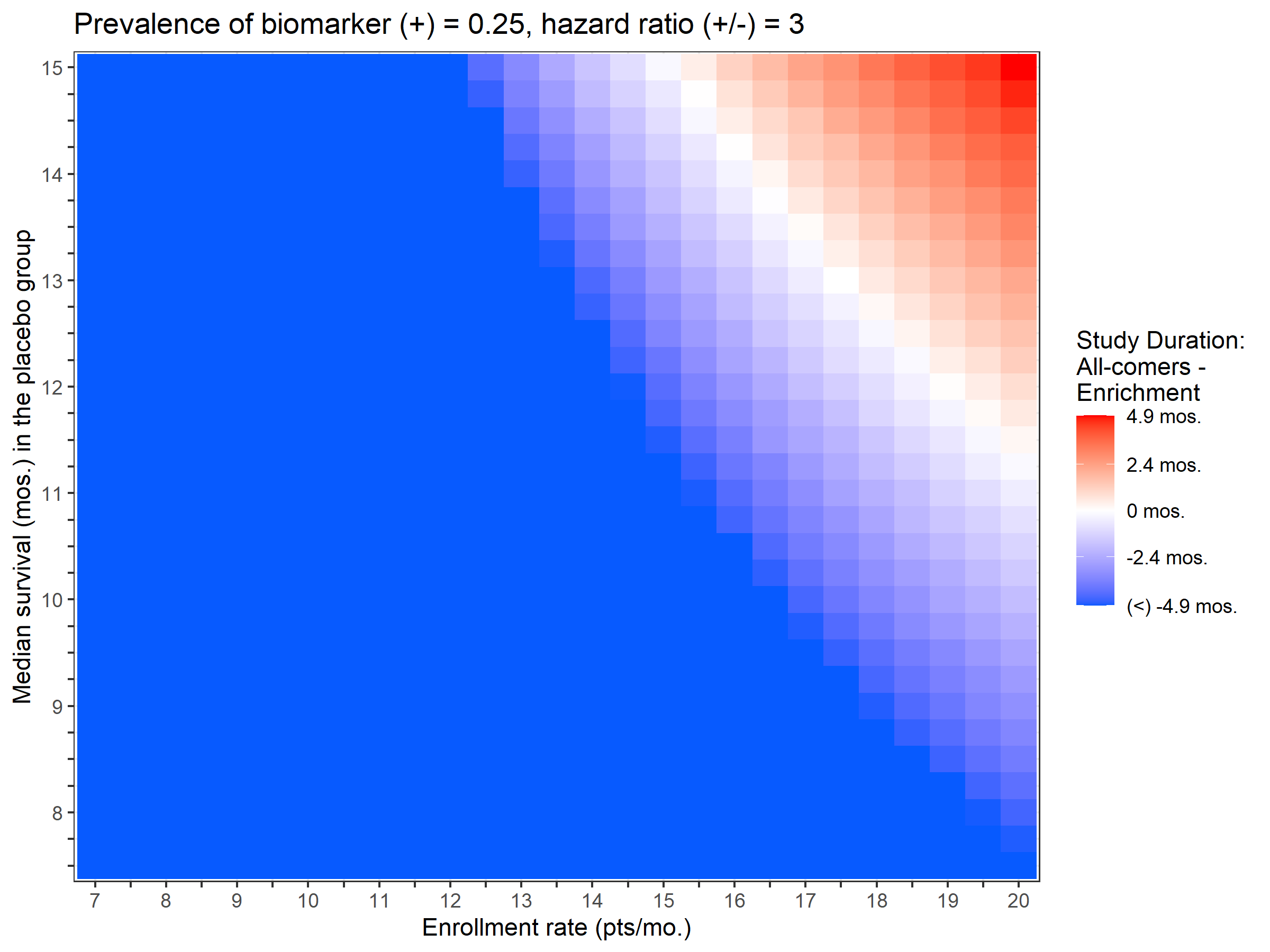}
\includegraphics[width=5.3cm, height=4cm]{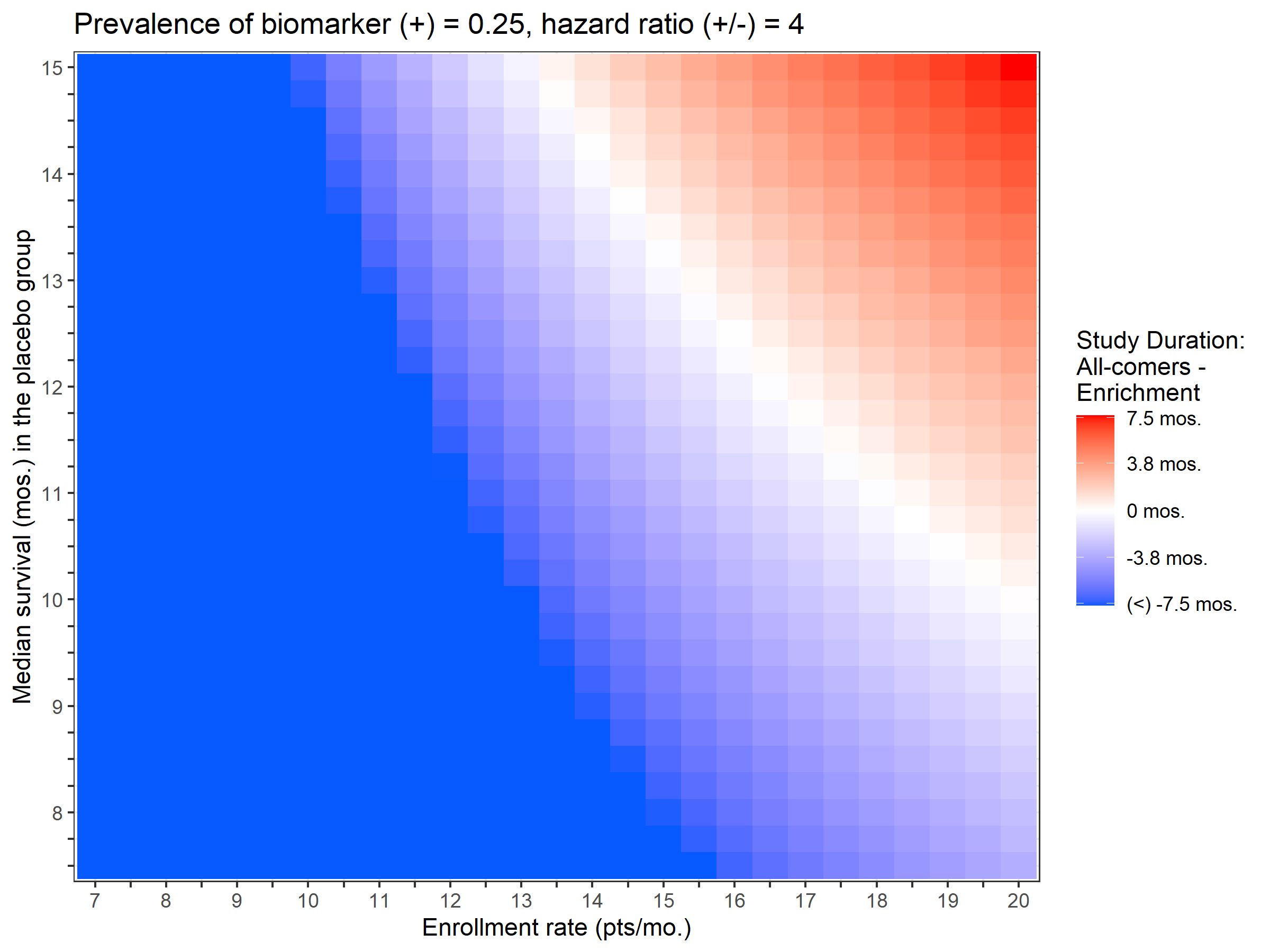}\\
\includegraphics[width=5.3cm, height=4cm]{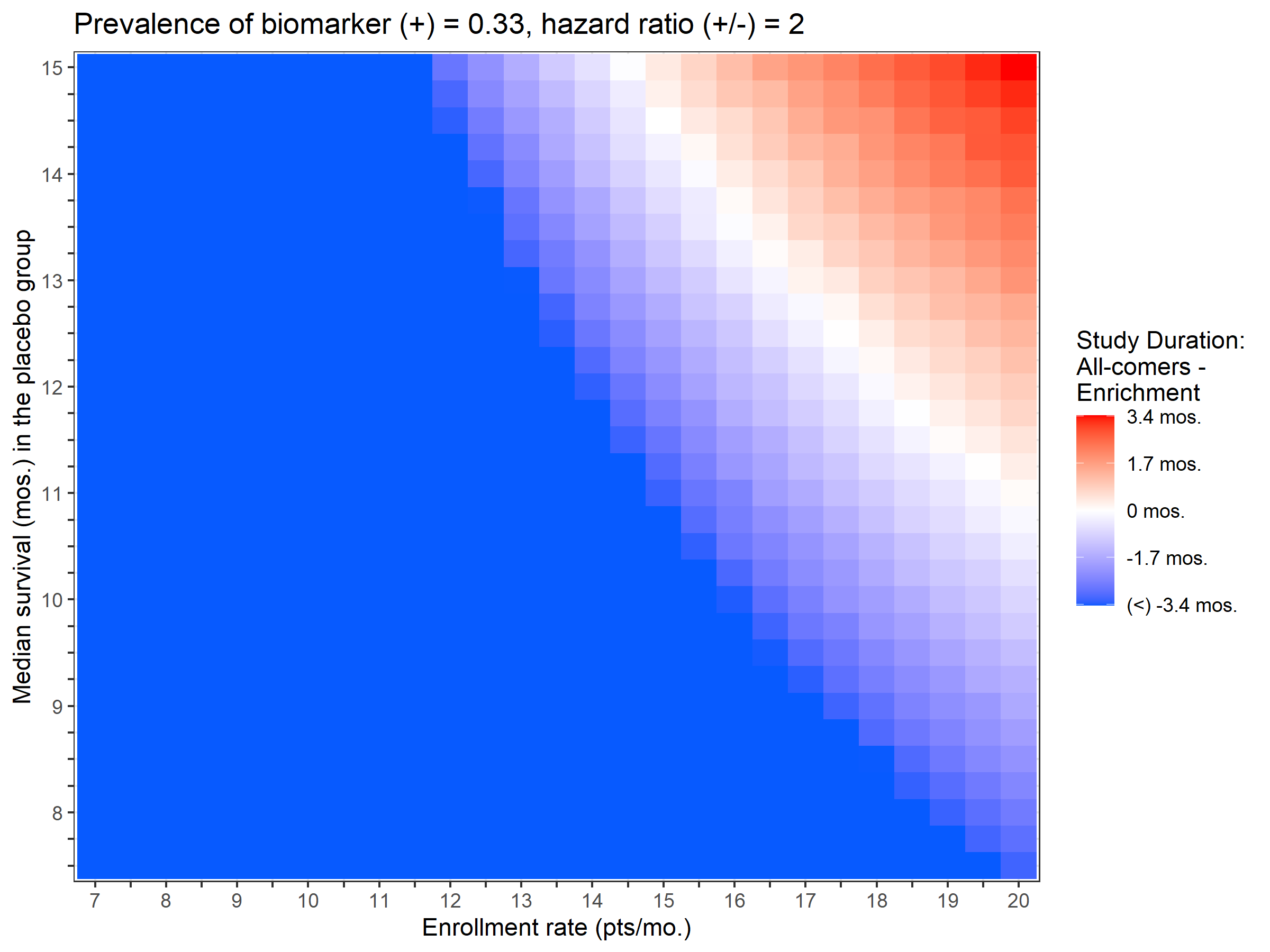}
\includegraphics[width=5.3cm, height=4cm]{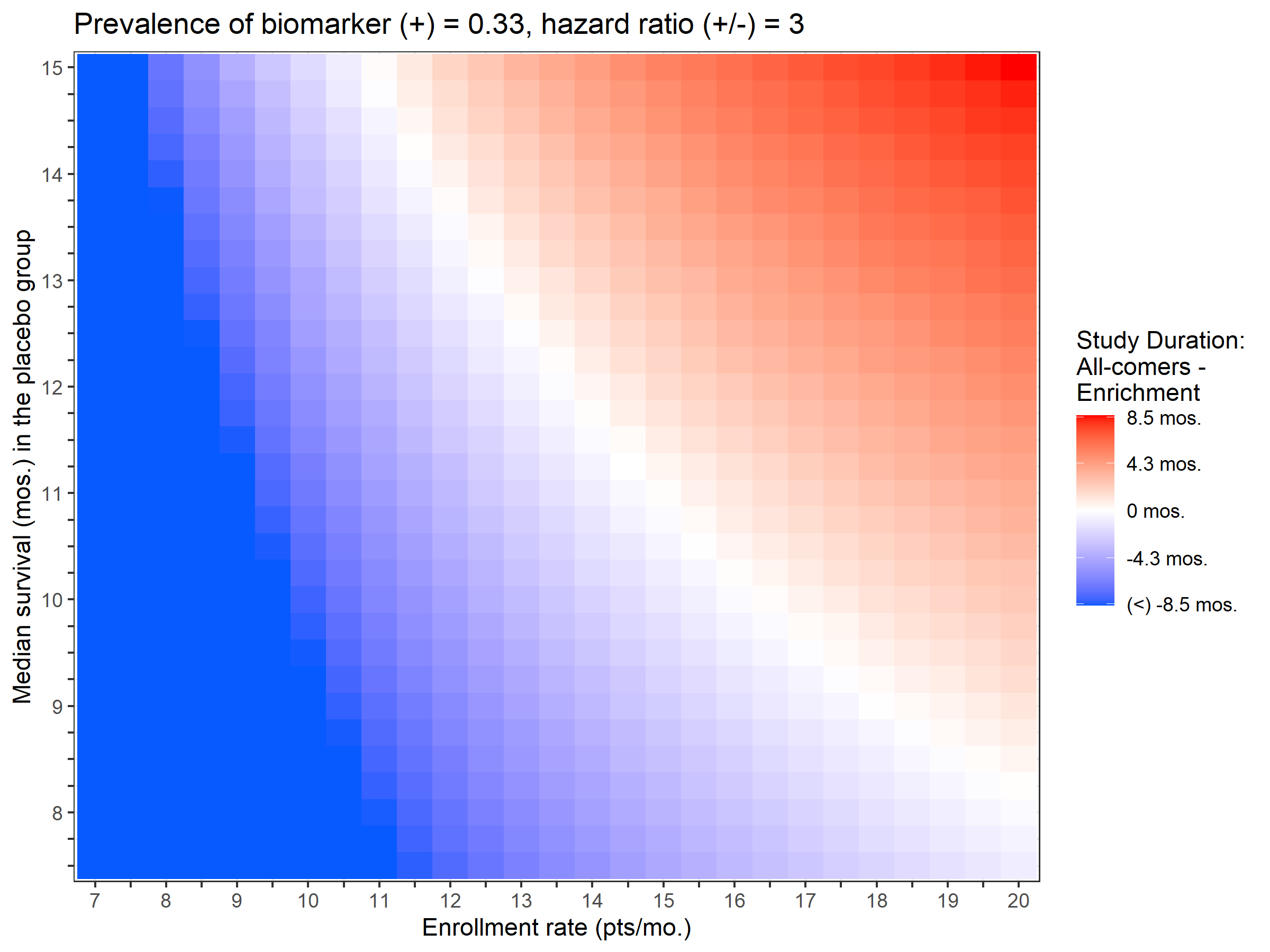}
\includegraphics[width=5.3cm, height=4cm]{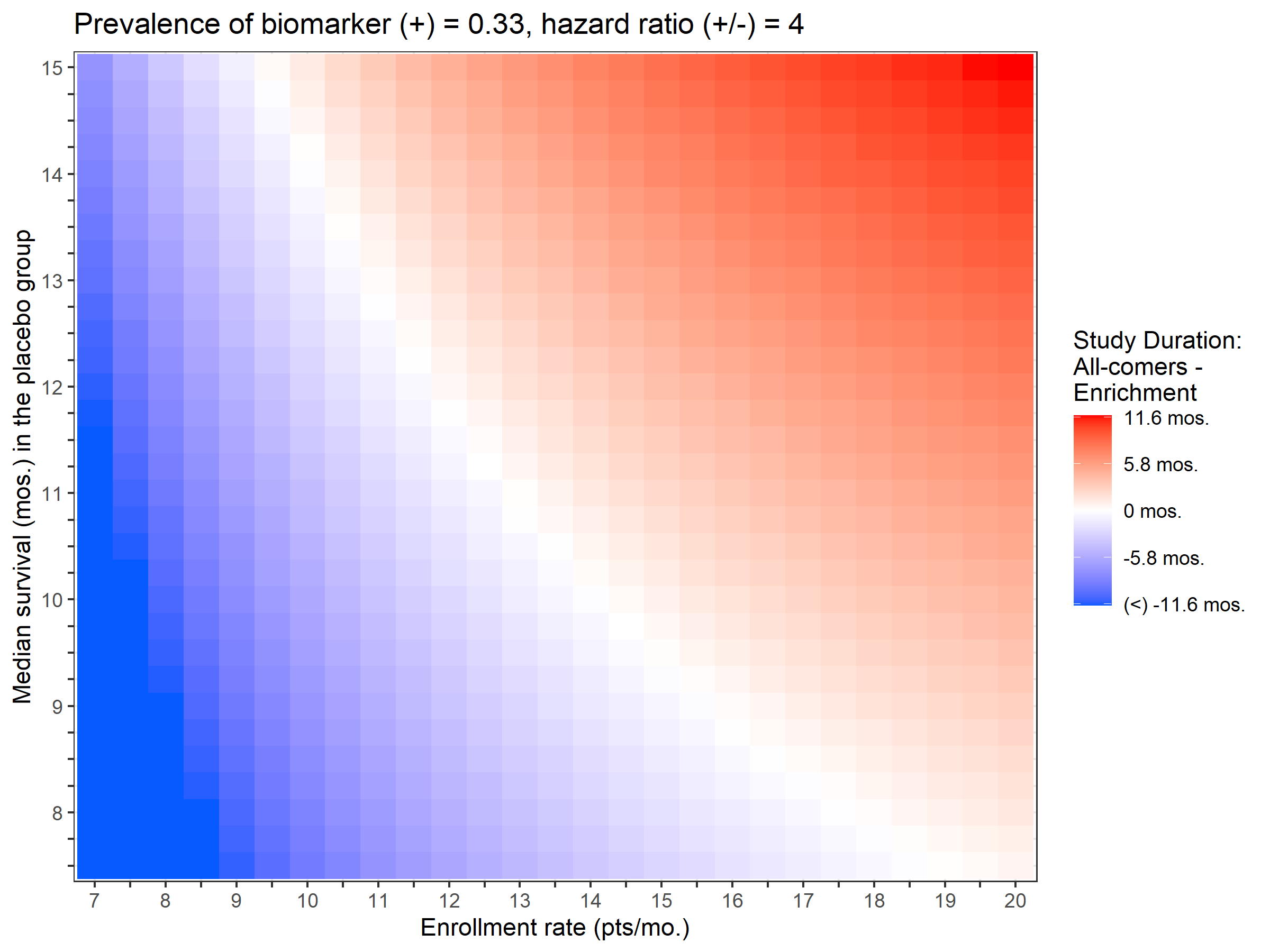}\\
\includegraphics[width=5.3cm, height=4cm]{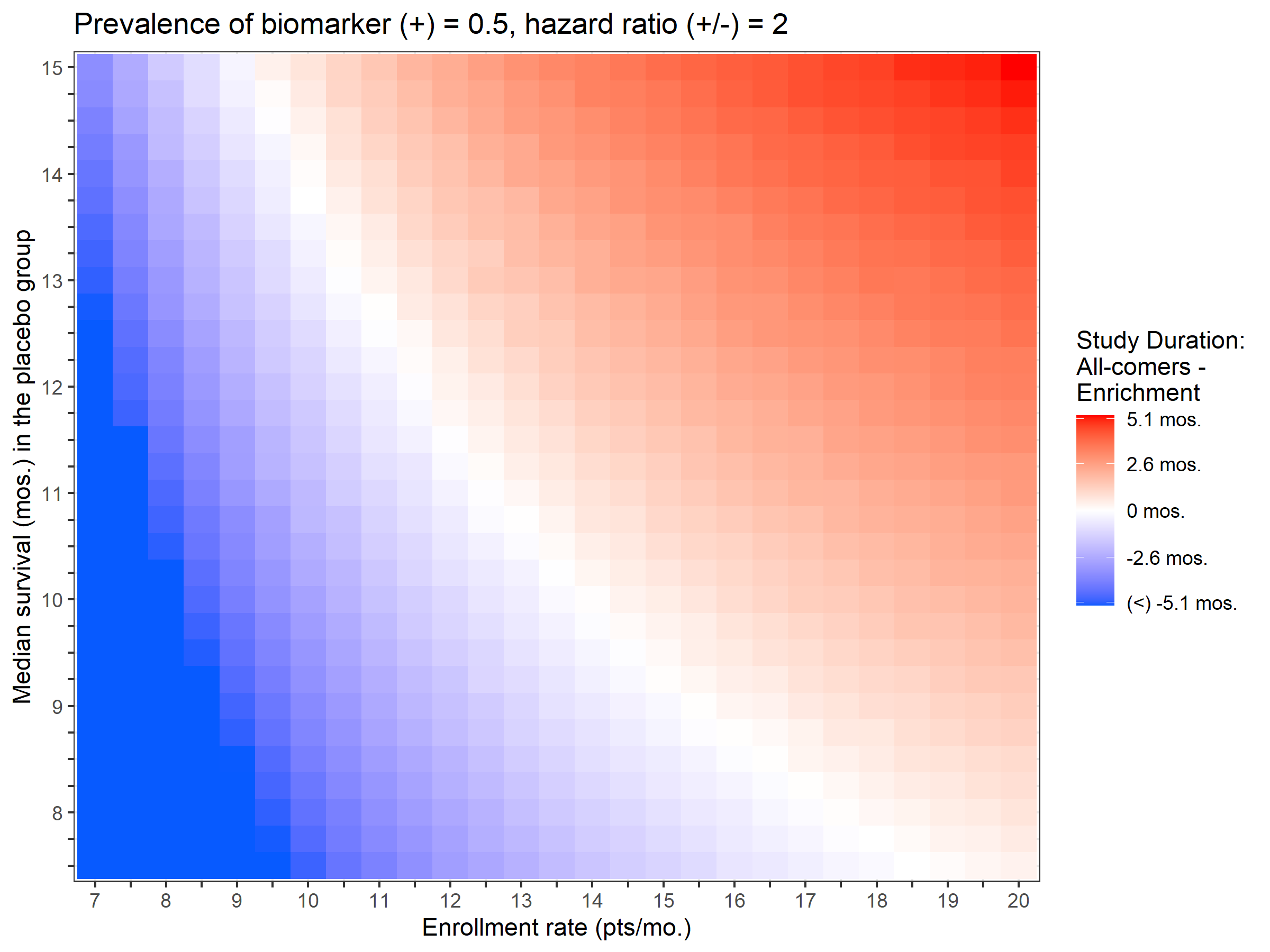}
\includegraphics[width=5.3cm, height=4cm]{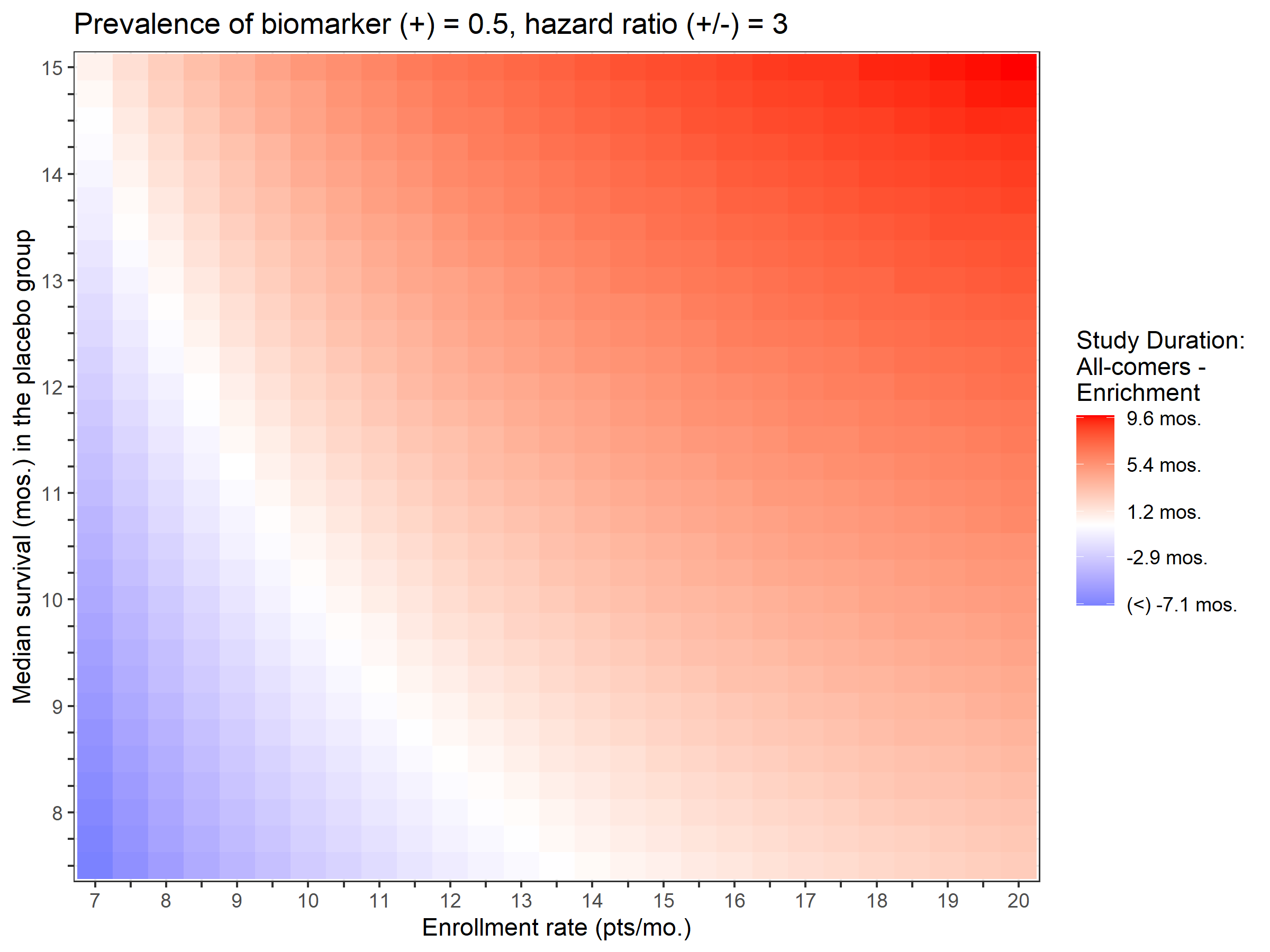}
\includegraphics[width=5.3cm, height=4cm]{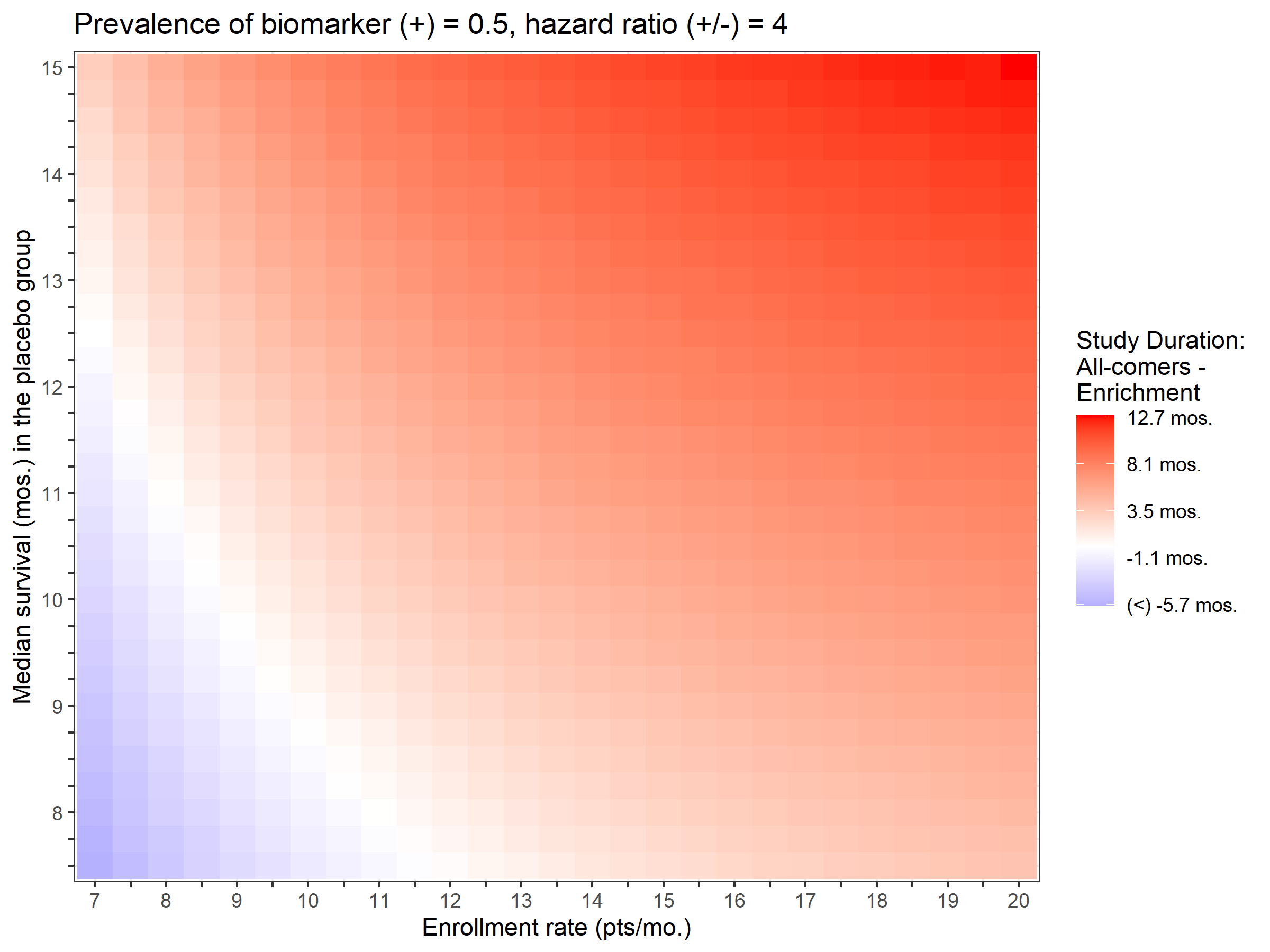}
\caption{The heatmap of the study duration difference between the all-comers design and the enrichment design.  Columns: the hazard ratio of the biomarker positive $=2, 3, 4$. Rows: the prevalence of the biomarker positive $=0.25, 0.33, 0.5$. within each panel, x-axis: the enrollment rate of the all-comers trial; y-axis: median survival time of the placebo group.}
\label{fig:supp1}
\end{figure}

\end{document}